\documentclass[prd,aps,twocolumn,floats,floatfix]{revtex4}

\usepackage{graphicx}
\usepackage{amsmath,amssymb}
\usepackage{physics}
\usepackage{subfigure}
\usepackage{appendix}
\usepackage{xcolor}
\usepackage{hyperref}
\usepackage{booktabs} 
\usepackage{longtable}

\begin{document}

\title{Cooling binary neutron star remnants via nucleon-nucleon-axion bremsstrahlung}

\author{
Tim Dietrich$^1$, 
Katy Clough$^2$
}

\affiliation{$^1$ Nikhef, Science Park 105, 1098 XG Amsterdam, The Netherlands}
\affiliation{$^2$ Astrophysics, University of Oxford, DWB, Keble Road, Oxford OX1 3RH, UK}

\date{\today}

\begin{abstract}
The QCD axion is a hypothetical particle motivated by the Strong CP problem of particle physics. 
One of the primary ways in which its existence can be inferred is via its function as an additional 
cooling channel in stars, with some of the strongest constraints coming from the supernova observation SN1987A. 
Multimessenger observations of binary neutron star mergers 
(such as those of GW170817, AT2017gfo, and GRB170817A) may
provide another scenario in which such constraints could be obtained. 
In particular, the axion could potentially alter the lifetime, the ejection of material, 
and the emitted gravitational wave signal of the postmerger remnant. 
In this article, we perform numerical relativity simulations of a binary neutron star merger, 
including a phenomenological description of the nucleon-nucleon-axion bremsstrahlung 
to quantify the effects of such a cooling channel on the dynamical evolution. 
While our simulations show a difference in the temperature profile of the merger remnant, 
the imprint of the axion via nucleon-nucleon-axion bremsstrahlung 
on the emitted gravitational wave signal and the ejecta mass is too small to improve 
constraints on the axion mass with current or future planned detectors. 
Whilst we consider a limited number of cases, and a simplified cooling model, 
these broadly represent the ``best case'' scenario, thus, a more thorough 
investigation is unlikely to change the conclusions, at least for this 
particular interaction channel.
\end{abstract}

\maketitle

\section{Introduction}
\label{sec:introduction}

Binary neutron star (BNS) mergers are among the most energetic events in our Universe. 
As such they provide a natural laboratory for investigating particle physics at high energies. 
The combined observation of gravitational waves (GWs) from a BNS merger 
by advanced LIGO and advanced VIRGO, GW170817, 
e.g.~\cite{TheLIGOScientific:2017qsa,Abbott:2018exr,Abbott:2018wiz,LIGOScientific:2018mvr}, 
along with the electromagnetic counterparts, 
AT2017gfo and GRB170817A, e.g.~\cite{Monitor:2017mdv,GBM:2017lvd,Coulter:2017wya}, 
were the first in what promises to be a catalog of future events, 
with ever increasing numbers of detections, and levels of accuracy, 
e.g.,~\cite{Abbott:2016ymx,Chruslinska:2017odi}.

\begin{figure}[t]
 \centering
 \includegraphics[width=\columnwidth]{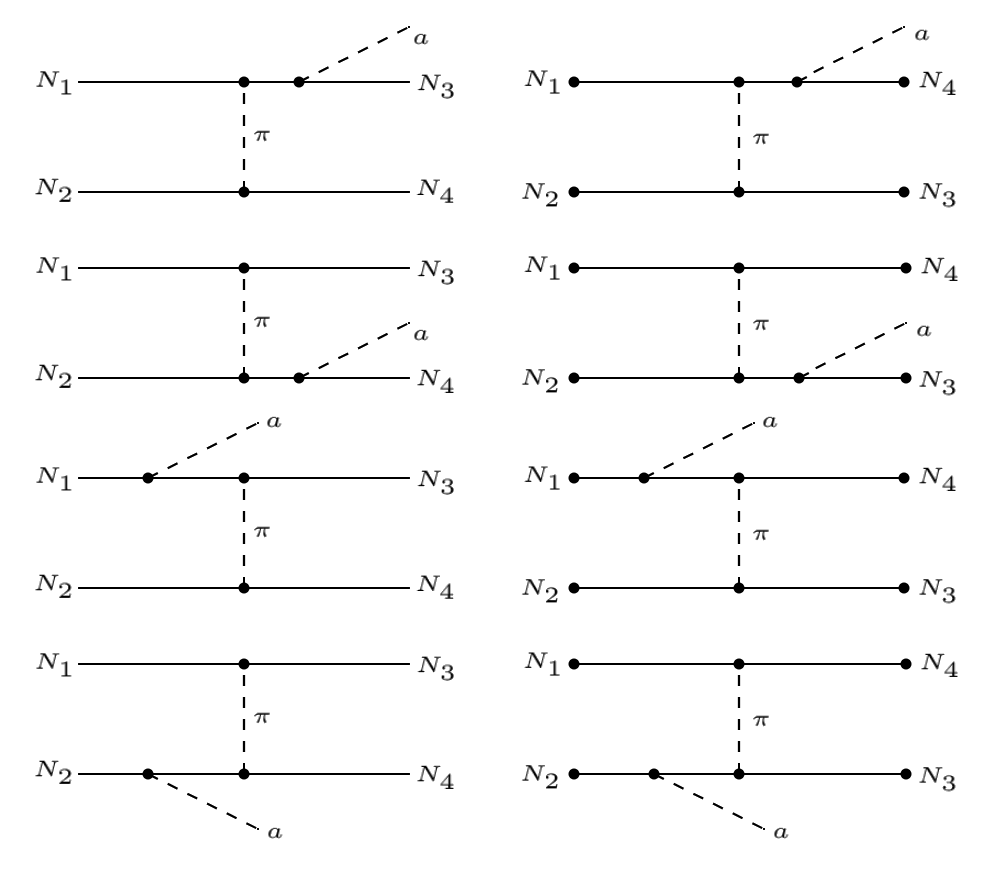}
 \caption{Direct and exchange diagrams corresponding to the 
 nucleon-nucleon-axion bremsstrahlung, see e.g.~\cite{Brinkmann:1988vi}.}
 \label{fig:feynman}
\end{figure}

In this paper we explore the use of BNSs for testing the existence of the axion, 
a hypothetical light pseudoscalar particle with sub eV mass. 
The original motivation for axions was to solve the strong CP problem in QCD, 
by providing a dynamical mechanism to explain the smallness of potential 
CP violating terms in the Standard Model Lagrangian \cite{Peccei:1977hh, Vafa:1984xg}
(see \cite{Hook:2018dlk} for a review). 
For this QCD axion, the axion decay constant $f_a$ (also the Peccei-Quinn energy scale for axions)
is related directly to the axion mass $m_a$ as
\begin{equation}
	m_a = \left( \frac{10^{6} {\rm GeV}}{f_a} \right) ~ 6 {\rm eV} ~ , \label{eq:axion}
\end{equation}
and, subject to some model dependence, calculations of the coupling of axions to photons 
$g_{a\gamma\gamma}$ and axions to nucleons $g_{a n n}$ can also be made where 
$g_{a\gamma\gamma} \sim 1/f_a$ and so higher masses correspond to stronger couplings. 
A wider class of axion like particles (ALPs) covers (pseudo)scalar particles arising from, 
for example, string theory, for which the couplings to standard model matter are 
unconstrained and may even be zero \cite{Svrcek:2006yi, Kim:1986ax}. 
In this work, we focus on the former, more predictive case, 
but our work is relevant to the production of any light particles with a similar coupling to nucleons. 
Another motivation for axions is as a dark matter candidate 
\cite{Preskill:1982cy, Abbott:1982af, Dine:1982ah} (see \cite{Marsh:2015xka} for a review), 
but this is not a required assumption for our work, 
which relies only on axion production by nucleon-nucleon-axion bremsstrahlung (Fig.~\ref{fig:feynman})
in the remnant, and not on an astrophysical background of axions. 

We build on a long history of astrophysical tests for axions 
(e.g. \cite{Dicus:1978fp, Vysotsky:1978dc, Raffelt:1999tx, Raffelt:1987np, Schlattl:1998fz, Isern:2003xj}, 
see \cite{Raffelt:2006cw} for a review),  
in which one uses the fact that production of axions in a dense plasma would change 
the behaviour and evolution of stars. Several such tests constrain the axion mass to be below 1 eV. 
In addition, in a supernova such as SN1987A, the formation of axions from 
nucleon-nucleon-axion bremsstrahlung would potentially act as a cooling mechanism, 
reducing the duration of the neutrino burst \cite{Ellis:1987pk, Raffelt:1987yt, Turner:1987by, 
Mayle:1987as, Mayle:1989yx, Burrows:1988ah, Burrows:1990pk, Brinkmann:1988vi, Keil:1996ju}. 
In \cite{Burrows:1988ah, Burrows:1990pk}, using the observation of SN1987A neutrinos, 
this effect was found to rule out QCD axion masses between $10^{-3}$eV and $1$eV. 
The lower limit is set by the interaction strength becoming sufficient for the
production of enough axions to enhance cooling. The upper limit arises when the interaction strength is 
too high - the freestreaming of axions ceases, 
due to the mean free path of the axions being contained within the core, such that the axions 
no longer carry away energy faster than the neutrinos. 
More recent reanalysis, e.g. \cite{Chang:2018rso}, have provided order of magnitude adjustments to
these original bounds, by taking into account more accurate models, for example, 
for the spin dependence of the axion coupling, 
but there remains a level of model dependence in the results and ranges between 
$10^{-3}$eV and $10$eV are obtained \cite{Raffelt:2006cw}.

Motivated by this work, we note that the same cooling effect should apply in a BNS merger, 
where the neutron star matter is rapidly heated during the collision.
After the merger, the remnant is not only supported against collapse by the intrinsic rotation, 
but also by temperature gradients.
An additional cooling mechanism should lead to changes 
in the collapse time to a black hole (BH), e.g.,~\cite{Paschalidis:2012ff,Gill:2019bvq}. 
Whilst this part of the GW signal was not observed for GW170817~\cite{Abbott:2017dke,Abbott:2018wiz}, 
one expects that with increasing signal to noise ratios (SNRs), 
future observations will constrain the time to collapse better, 
giving a possible measure of the cooling timescale. 
Furthermore, observations of electromagnetic counterparts may give other distinctive hints of 
the presence of additional cooling pathways. 
Note that neutrino emission is one known channel for BNS 
cooling, e.g., \cite{Rosswog:2003rv,Dessart:2008zd,Paschalidis:2012ff}, 
which is likely to dominate in the BNS case, and
although a model for this is not yet fully agreed upon in numerical simulations 
(see, e.g. \cite{Sekiguchi:2011zd,Galeazzi:2013mia,Foucart:2014nda,Neilsen:2014hha,
Palenzuela:2015dqa,Lehner:2016lxy,Foucart:2018gis}), 
models will hopefully converge in the future. 
In this work we neglect the effects of neutrino cooling, to test the impact of axion 
cooling independently. Consequently, our constraints can be seen as the most optimistic 
analysis. We investigate the range between $10^{-3}$ eV and 1 eV, 
which overlaps with the SN1987A range, and so could potentially reinforce these 
constraints using an independent type of event.

We note that in addition to nucleon-nucleon-axion bremsstrahlung
there are a number of other ways in which axions may change the BNS coalescence. 
In particular, the imprint of axions or ALPs on the BNS GW signals observed by LIGO and Virgo 
(see e.g. \cite{Brito:2017zvb, Blas:2016ddr, Rozner:2019gba, Hook:2017psm, Huang:2018pbu}).
These effects generally rely on the Compton wavelength of the axion particle being of an 
astrophysical scale - i.e. similar to the size of the binary or individual stars, 
such that the wave-like nature of the scalar condensate leads to effects 
such as superradiance \cite{Brito:2015oca}, orbital resonances, 
or a distinctive pattern in the density in or around the stars.
For the axion masses considered here, 
these effects would not be relevant since the Compton wavelengths are sub mm, 
and the behavior is well described by the particle picture. 
Another possible effect is the imprint of accumulated DM
in the NS cores (see e.g. \cite{Brito:2015yfh,Ellis:2017jgp,Bezares:2019jcb}). 
As noted above, we will not assume that the axion constitutes the DM, 
and so do not consider these effects in our simulations.\\

In this paper, we present a first step in modelling and quantifying the effect of the 
nucleon-nucleon-axion bremsstrahlung and structure the article as follows: 
First, we discuss the introduced phenomenological cooling scheme in Sec.~\ref{sec:methods}
and the numerical setup and simulated configurations in Sec.~\ref{sec:simulations}.
Results regarding the dynamical evolution, the GW signal, and the mass ejection in our simulations
are presented in Ref.~\ref{sec:results}, we conclude and summarize in Sec.~\ref{sec:summary}.
Throughout this work we employ geometric units $c=G=M_\odot=1$ if not otherwise stated. 
This determines the units of all our results, in particular, those shown in the individual 
figures. For additional details and the conversion into standard units, 
we refer to Appendix~\ref{app:units}. 

\section{Axion cooling}
\label{sec:methods}

\subsection{Nucleon-Nucleon-Axion Bremsstrahlung}
\label{sec:NNAB}

For a simple model of axion production, we follow the treatment in 
Brinkmann et al.~\cite{Brinkmann:1988vi}, which is summarized below for completeness.

In this model, the degenerate and non-degenerate energy emission is given by 
\begin{eqnarray}
\dot{\epsilon}_D & = & 5.3 \times 10^{44} [{\rm erg\ cm^{-3} s^{-1}}] f^4 g^2 (X_N \rho_{14})^{1/3} T^6_{\rm MeV} \nonumber \\ 
\dot{\epsilon}_{ND} & = & 1.1 \times 10^{47} [{\rm erg\ cm^{-3} s^{-1}}] f^4 g^2 (X_N \rho_{14})^{2} T^{3.5}_{\rm MeV}\nonumber 
\end{eqnarray}
with $X \sim 1$ being the mass fraction of nucleons, $f \sim 1$ the pion-nucleon coupling, 
$\rho_{14} = \rho / 10^{14} ~ \rm g ~ cm^{-3}$ the energy density of the remnant, 
$T_{\rm MeV} = T / 1 ~ \rm MeV$ the temperature, and $g$ the axion-nucleon coupling.
The value of $g$ is a function of the nucleon mass $m_n = 0.94$ GeV and $f_a$, 
and so can be expressed in terms of the axion mass $m_a$ using Eq.~\eqref{eq:axion}, as
\begin{equation}
g = m_n / f_a ~  \approx 1.516 \times10^{-7} \left( \frac{m_a}{\rm 1 eV} \right) ~ .
\end{equation}

Transforming these expressions into geometric units suitable for our simulations (see Appendix~\ref{app:units}), 
we arrive at: 
\begin{eqnarray}
\dot{\epsilon}_D & = & 1.98\times 10^{-12}\ \left( \frac{m_a}{\rm 1 eV} \right)^2\ \rho^{1/3}\ T^6_{\rm MeV}, \nonumber \\ 
\dot{\epsilon}_{ND} & = & 8.56  \times 10^{-4}\ \left( \frac{m_a}{\rm 1 eV} \right)^2\ \rho^{2}\ T^{3.5}_{\rm MeV}\nonumber .
\end{eqnarray}

As shown in the numerical simulations in Ref.~\cite{Brinkmann:1988vi} 
the degenerate approximation is 
accurate for small temperatures, whilst the non-degenerate emission is accurate for high temperatures. 
We enforce this by implementing
\begin{equation}
 \dot{\epsilon} = \min (\dot{\epsilon}_D, \dot{\epsilon}_{ND}). \label{eq:minepsilon}
\end{equation}

We note that in the case of our higher masses - above $m_a \sim 0.01 eV$, 
the effect of reabsorption of the axions should be taken into account \cite{Burrows:1990pk}, 
as at this point their mean free path falls below one NS radius and some of the 
axions will be thus reabsorbed in the remnant. 
Above 1 eV virtually no axions would escape and cooling from this channel would cease. 
We have not taken these effects into account and as such our simulations represent 
a ``best case scenario'' for the potential level of cooling. 
As discussed below, we nevertheless find that the effect is small, 
and so such corrections would only further reduce the potential constraints which can be obtained.

\subsection{A covariant cooling scheme}

In this article, we follow the treatment of cooling mechanisms in Refs.~\cite{Paschalidis:2011ez,Paschalidis:2012ff}.
A covariant extension of the general relativistic hydrodynamics (GRHD) equations is given 
to incorporate additional cooling channels, e.g., neutrino radiation as
considered in~\cite{Paschalidis:2012ff}. 
Here we will briefly review this formalism and describe the modifications to 
incorporate nucleon-nucleon-axion bremsstrahlung. 
We introduce 
\begin{equation}
 \nabla_\alpha R^{\alpha \beta} = - G^{\beta}
\end{equation}
with the radiation stress-energy tensor $R^{\alpha \beta}$
and the radiation four-force density $ G^{\alpha}$.
The energy-momentum conservation equation reads
\begin{equation}
 \nabla_\alpha (T^{\alpha \beta} + R^{\alpha \beta} ) = 0 
\end{equation}
or alternatively 
\begin{equation}
 \nabla_\alpha T^{\alpha \beta} = G^\beta. 
\end{equation}
Assuming a perfect fluid stress-energy tensor and projecting this equation with respect to the fluid four-velocity,
we obtain, with the help of the continuity equation,
\begin{equation}
 u^\alpha \nabla_\alpha \epsilon  = \frac{\epsilon + p}{\rho_0} u^\alpha \nabla_\alpha \rho_0 - u^\alpha G_\alpha ~.
\end{equation}
We relate the thermal part of $\epsilon$ to the temperature via
\begin{equation}
 \epsilon_{\rm th} = \frac{3 k T}{2 m_u} + \frac{11 a_r T^4}{4 \rho}
\end{equation}
with the Boltzmann constant $k$, the atomic mass unit $m_u$, the radiation-density constant $a_r$. \\

Under the assumption (discussed in Sec. \ref{sec:NNAB} above) that there is no absorption or scattering, we set $G^\alpha = - u^\alpha \Lambda$
and $u_\alpha G^\alpha = \Lambda$.
We obtain the modified GRHD equations as 
\begin{equation}
 \partial_t S_i + \partial_j (\alpha \sqrt{\gamma} T^j_i) = \frac{1}{2} \alpha \sqrt{\gamma} T^{\alpha \beta} 
 \partial_i g_{\alpha \beta } - \alpha \sqrt{\gamma} u_i \Lambda 
\end{equation}
and 
\begin{equation}
 \partial_t \tau + \partial_i (\alpha^2 \sqrt{\gamma} T^{0i} - D_i v^i) = s - \alpha^2 \sqrt{\gamma} u^0 \Lambda.
\end{equation}

To connect the derived cooling scheme to the axion bremsstrahlung, Eq.~\eqref{eq:minepsilon}, 
we set 
\begin{equation}
 \Lambda = \dot{\epsilon},
\end{equation}
which closes our system of equations.

\begin{figure*}[t]
 \centering
 \includegraphics[width=1.9\columnwidth]{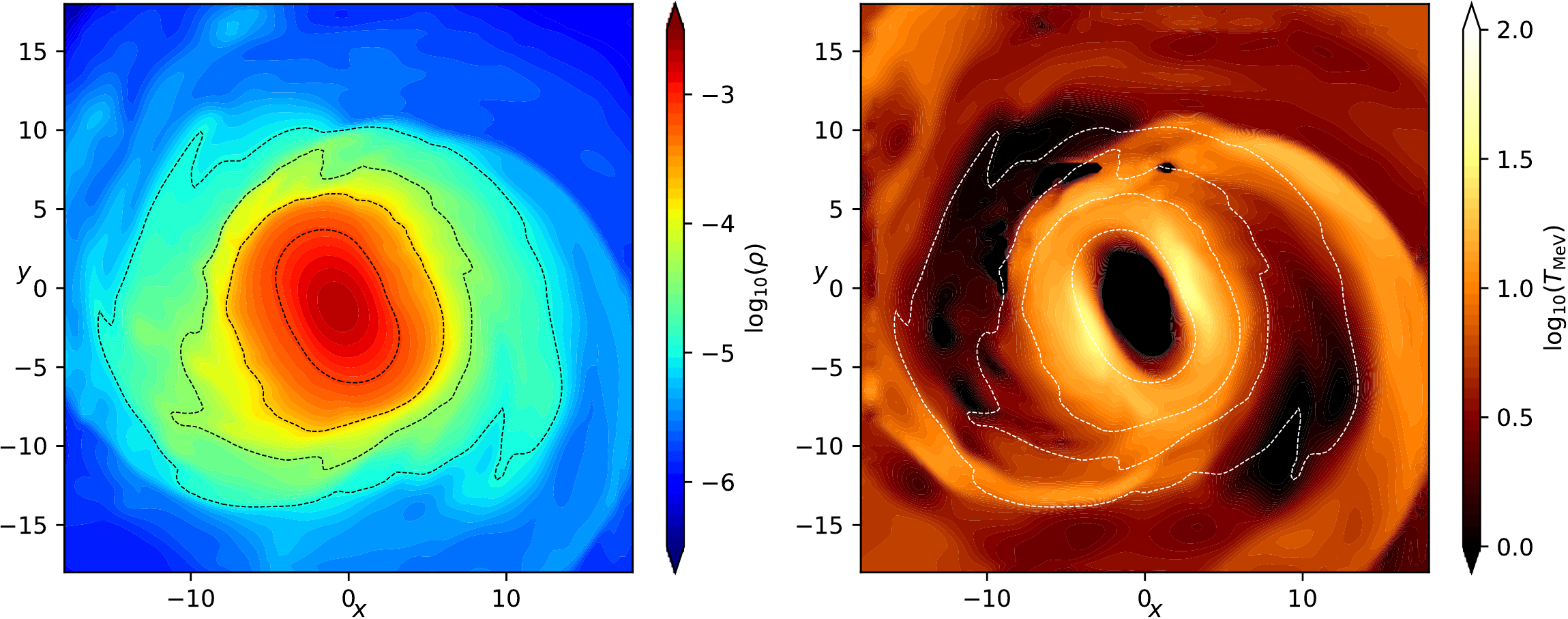}
 \includegraphics[width=1.9\columnwidth]{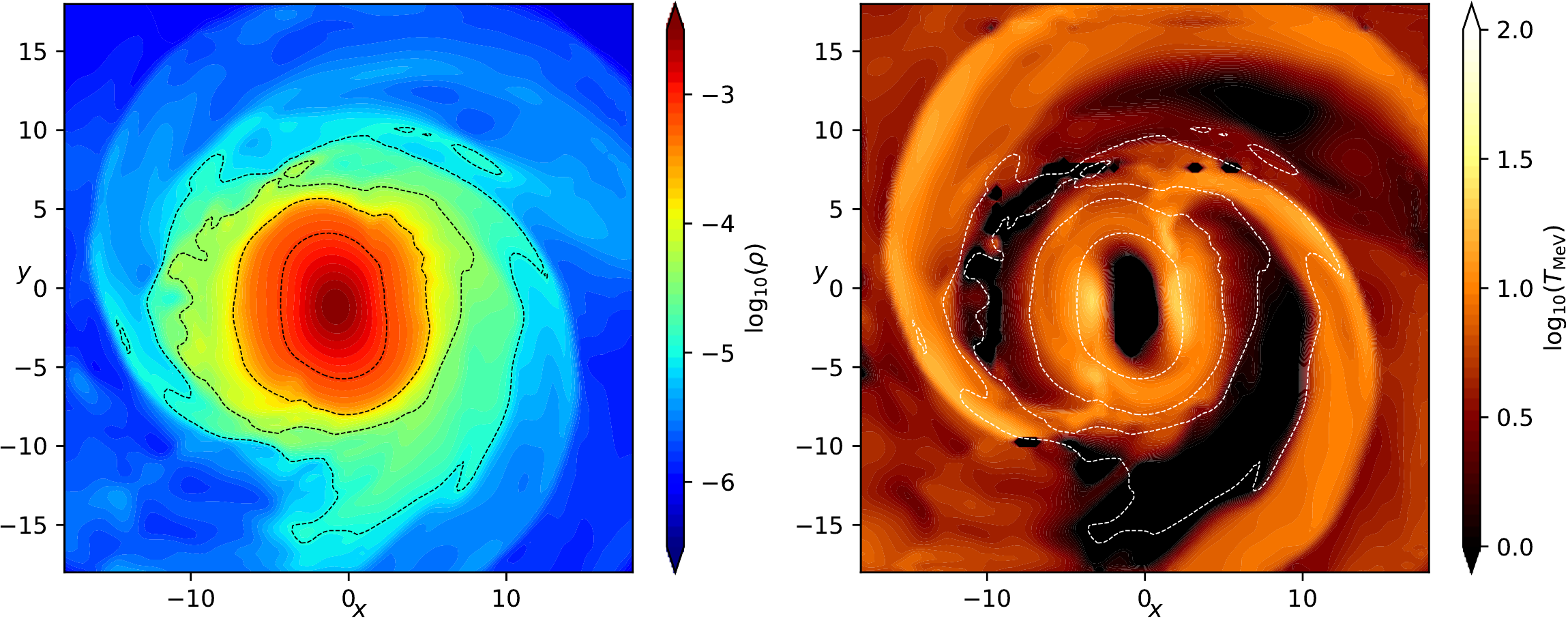} 
 \includegraphics[width=1.9\columnwidth]{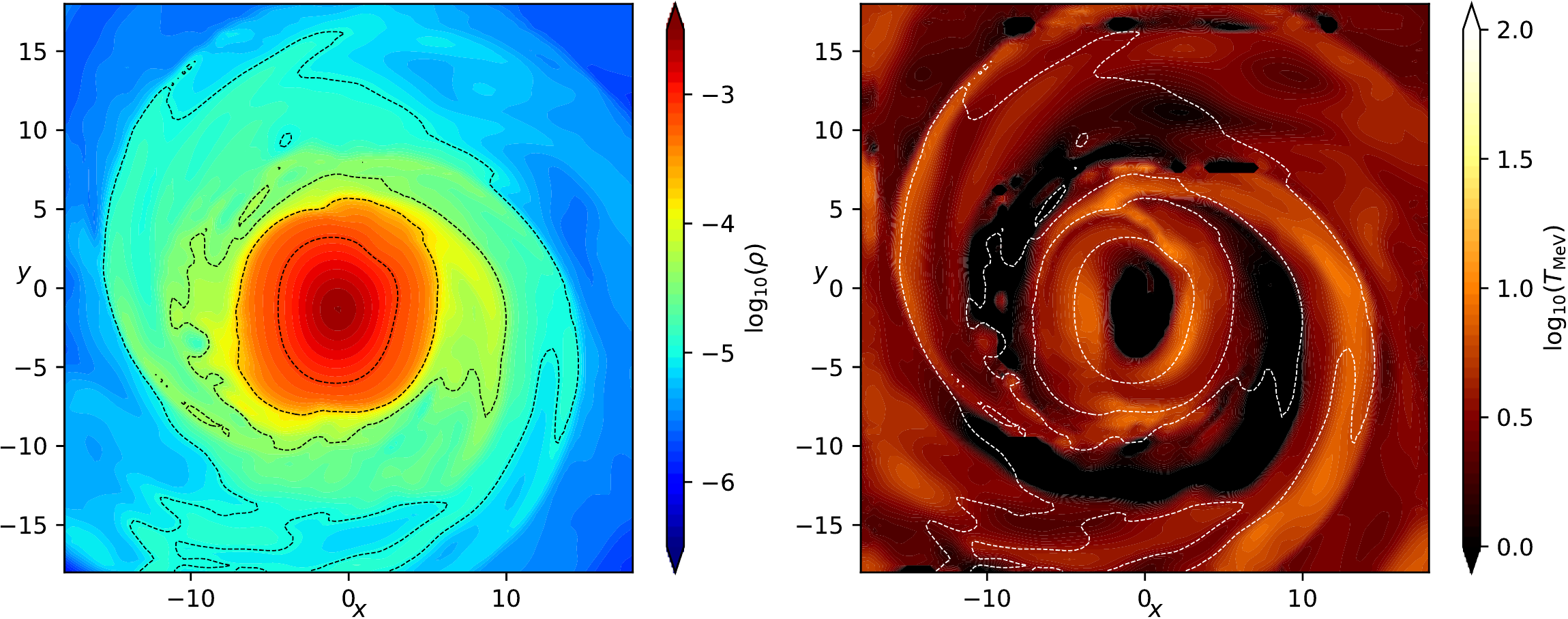}
 \caption{Density profile (left) and temperature profile (right) for the configuration Case-1 
 at a time $t=3000$. The individual rows refer to $m_{a,eV}=0,0.1,1$ from top to bottom. 
 Clearly visible is a more spherical density profile and a smaller temperature in the presence of 
 nucleon-nucleon-axion bremsstrahlung. 
 All individual panels employ the same plot range to allow a better comparison.}
 \label{fig:HM4235}
\end{figure*}

\section{Simulated configurations and numerical methods}
\label{sec:simulations}

In this article, we are studying equal-mass, irrotational BNS configurations for 
four different total masses, cf.\ Tab.~\ref{tab:config}. 
All setups employ a piecewise polytropic representation of the 
ALF2 EOS~\cite{Douchin:2001sv,Read:2008iy}. 
This EOS has been picked since it is in agreement with current constraints on the supranuclear EOS 
derived from GW170817, AT2017gfo, GRB170817A, e.g.~\cite{Radice:2018ozg,Coughlin:2018fis}~\footnote{We 
note that the most recent analysis of GW170817~\cite{Capano:2019eae}, 
which has been published while our manuscript was in internal circulation, 
suggests neutron star radii slightly smaller than predicted by the ALF2 EOS.}. 
We model the imprint of different axion masses, by considering 
$m_{a} \in \{10^0, 10^{-1},10^{-2},10^{-3}, 0 \}$ eV, 
i.e., we perform simulations for four different axion masses, plus the null case. 

The initial configurations for our study are computed with the SGRID 
code~\cite{Tichy:2009yr,Tichy:2012rp,Dietrich:2015pxa} and evolved with 
BAM~\cite{Brugmann:2008zz,Thierfelder:2011yi,Dietrich:2015iva,Dietrich:2018bvi}. 
SGRID uses pseudospectral methods to compute spatial derivatives and solves 
the conformal thin sandwich equations~\cite{Wilson:1995uh,Wilson:1996ty,York:1998hy,Tichy:2016vmv}.
We do not employ any eccentricity reduction procedure since the focus of this work is on the 
postmerger evolution of the remnant and the inspiral is short - 
the merger occurs after 1-2 orbits. 

We use BAM for our dynamical simulations. The spacetime evolution follows the Z4c 
framework~\cite{Bernuzzi:2009ex,Hilditch:2012fp} combined with the 1+log and gamma-driver conditions 
for the evolution of the lapse and shift~\cite{Bona:1994a,Alcubierre:2002kk,vanMeter:2006vi}. 

For the evolution of the matter variables, we use high-resolution-shock-capturing 
methods~\cite{Thierfelder:2011yi} with primitive reconstruction 
(with the 5th order WENOZ method~\cite{Borges:2008a}) and the Local-Lax-Friedrichs numerical
flux scheme. 
For the nested Cartesian boxes, we employ a total of 7 refinement levels following a 2:1 refinement strategy. 
The three outermost levels are non-moving with a size of $192$ points in each dimension. 
The inner levels follow the motion of the NSs, each star is covered with approximately $96$ 
points in each dimension leading to a grid spacing of $0.177$. 
We enforce bitant symmetry to reduce the computational costs and the memory footprint of our simulations.
A detailed description of the numerical methods and convergence tests can 
be found in Refs.~\cite{Brugmann:2008zz,Thierfelder:2011yi,Bernuzzi:2016pie,Dietrich:2018upm}. 
Further information about the constraint violations for our simulations 
can be found in Appendix \ref{app:constraints}.

\begin{table}[t]
  \centering    
  \caption{Configuration details for the configurations studied.
  The columns refer to the name of the setup, 
  the gravitational mass of the individual stars, 
  the baryonic mass of the individual stars, 
  the ADM-mass, and the initial orbital frequency.}
  \begin{tabular}{c|cccc}        
    \hline
    name & $M^{A,B}$ & $M_b^{A,B}$ & $M_{\rm ADM}$ &$\Omega$ \\
    \hline
    Case-1   &  1.365 &  1.506    & 2.701 & 0.01093 \\
    Case-2   &  1.332 &  1.463    & 2.630 & 0.01189 \\    
    Case-3   &  1.294 &  1.420    & 2.560 & 0.01174 \\
    Case-4   &  1.222 &  1.334    & 2.419 & 0.01141 \\ 
    \hline
  \end{tabular}
 \label{tab:config}
\end{table}

\section{Results}
\label{sec:results}

\subsection{Postmerger Dynamics}

While there are a number of possible effects of axions and ALPS on the BNS coalescence, 
we focus purely on the additional cooling channel provided by the nucleon-nucleon-axion bremsstrahlung. 
We therefore focus our discussion on the postmerger stage of the BNS coalescence, at times when the high temperatures
which favor the production of axions are present. 

Due to the additional cooling, 
the temperature and density profile of the remnant changes noticeably for different axion masses. 
As an example, we show for the Case-1 configuration the system at a time $t=3000$ in 
Fig.~\ref{fig:HM4235}. 
From top to bottom the individual columns refer to $m_{a}=0,0.1,1$ eV respectively, i.e., 
an increasing axion mass. 
From the density profile (left panels) one finds that at the same times, remnants which are cooled 
due to nucleon-nucleon-axion bremsstrahlung are almost spherically symmetric and their 
bar-mode is noticeably less pronounced. 
Considering the temperature evolution, the configuration without additional axion cooling 
has significantly higher temperatures in the spiral arms formed during and after the merger.  
These higher temperatures might allow for higher electron fractions 
within the ejected material, which (for a quantitative analysis) would require 
an inclusion and accurate modeling of neutrino effects. 
The central core has a low temperature for all studied systems. 

\begin{figure}[t]
 \centering
 \includegraphics[width=\columnwidth]{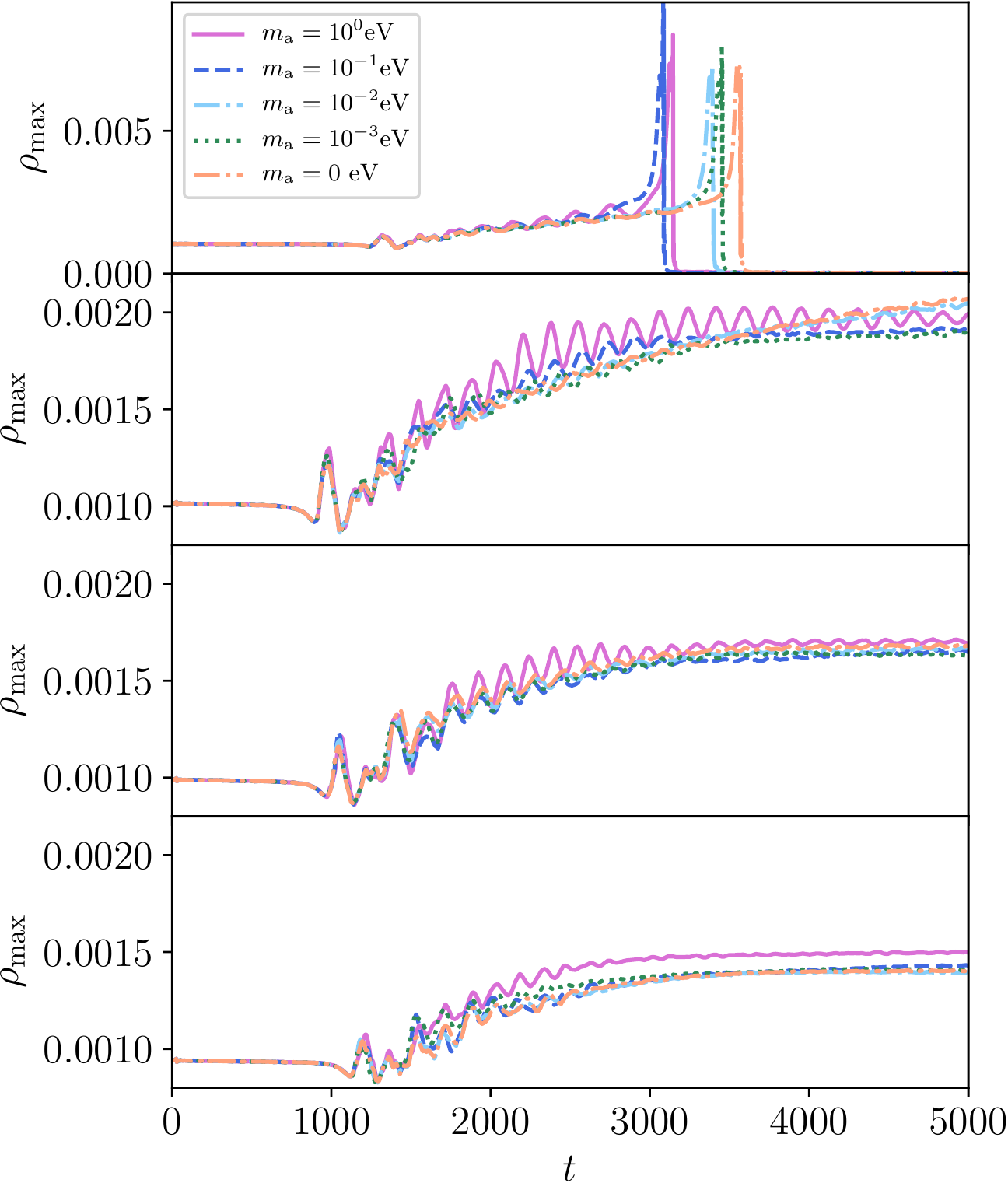}
 \caption{Maximum Density evolution for all four configurations:
 Case-1 (top), Case-2 (second panel), Case-3 (third panel), Case-4 (bottom panel). 
 Different colors and dashing corresponds to different axion masses. }
 \label{fig:rho_max}
\end{figure}

We show a comparison of the central density evolution for all four cases (Tab.~\ref{tab:config}) in Fig.~\ref{fig:rho_max}. 
We find that for the most massive system, the lifetime of the merger remnant 
depends on the axion cooling, where generally more cooling leads to an earlier collapse, 
we see that the lifetime is continuously decreasing for $m_{a}=0,10^{-3},10^{-2}\rm eV$, 
but that, somewhat surprisingly, $m_{a}=1\rm eV$ has a longer lifetime than $m_{a}=0.1\rm eV$. 
While this observation seems non-intuitive, the merger remnant evolution 
is complicated and determined by a number of factors. 
Most notably, one sees a clear oscillation of the maximum density throughout the entire evolution. 
This oscillation causes a non-monotonic evolution of the density and appears to be the reason 
for the longer lifetime of the $m_{a}=1\rm eV$ setup - causing a critical threshold to be crossed earlier in this case. 

Considering the three other cases (for which no BH forms during the numerical relativity simulation),
one generally finds that larger central densities are reached for configurations which incorporate
nucleon-nucleon-axion bremsstrahlung. The central density increases for larger axion masses. 
With decreasing axion masses the density evolution approaches the 
system without axion cooling, as expected. Even for an axion mass of $m_{a}=1\rm eV$ and 
for the system with the lowest total mass (Case-4)
the central density value is about $\sim 7\%$ larger than for the simulation excluding axion cooling. 
In addition, as for Case-1, we find that for the cases not forming a BH, the maximum density shows large oscillations. 
These oscillations are caused by the repulsion of the two colliding cores. 
For large axion masses, e.g., $m_{a}=1\rm eV$, these oscillations are larger than for 
smaller axion masses.

\subsection{GW emission}

\begin{figure}[t]
 \centering
 \includegraphics[width=\columnwidth]{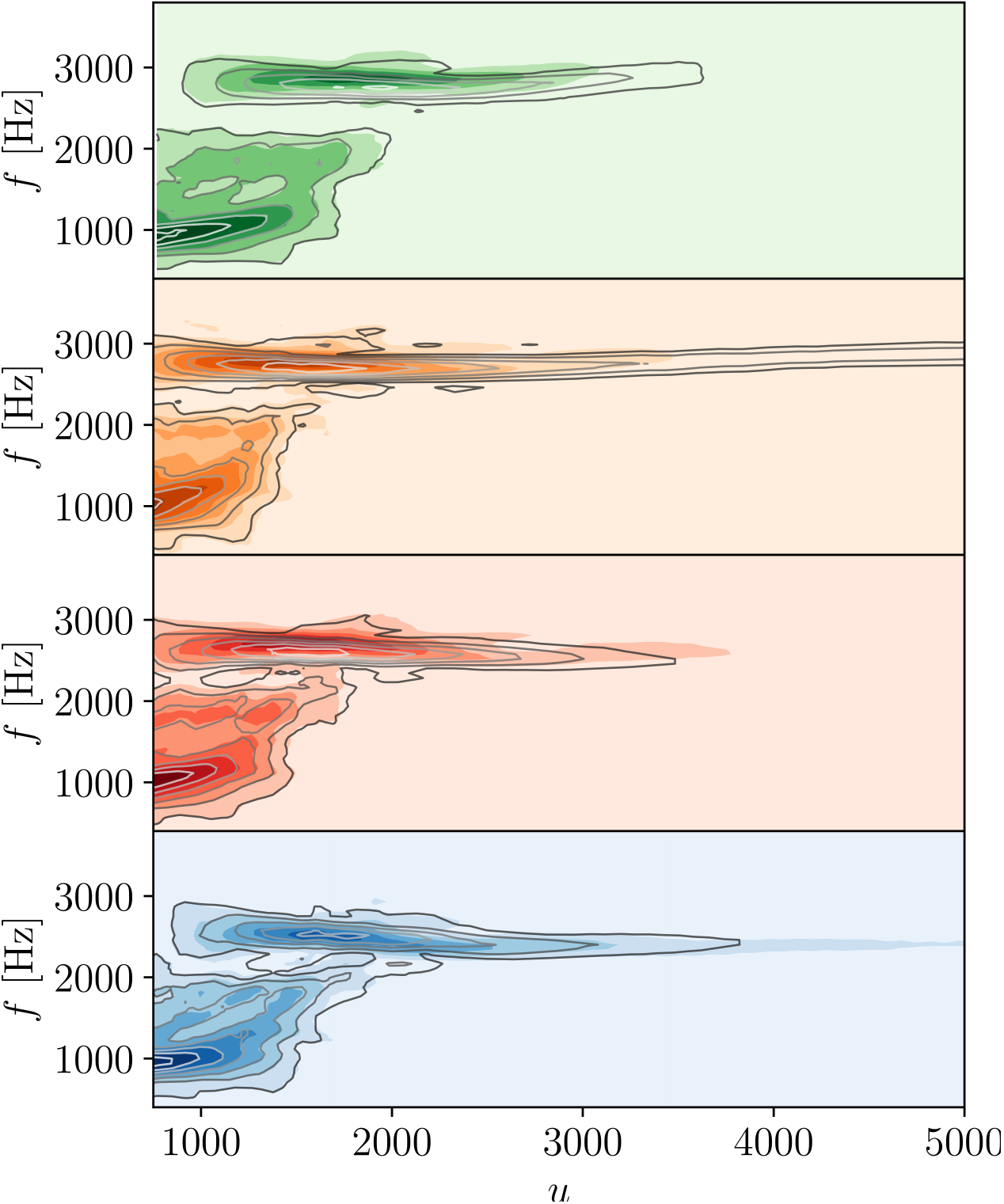}
 \caption{GW Spectrogram for the four studied cases:
 Case-1 (top), Case-2 (second panel), Case-3 (third panel), Case-4 (bottom panel). 
 Different colors and dashing corresponds to different axion masses.
 The colored contour plot refers to the simulations with an axion mass of $m_{a}=1\rm eV$, 
 the black contours refer to a simulation with an axion mass set to zero. 
 The spectrogram is computed by including all individual modes up to 
 $(\ell,m)=(4,4)$ and assuming an inclination of $\iota=\pi/2$.}
 \label{fig:GW_spectrum}
\end{figure}

As discussed, the presence of nucleon-nucleon-axion bremsstrahlung leads to a 
more spherical form of the merger remnant and thus potentially
faster BH formation. One might therefore expect a detectable imprint on the postmerger GW signal.  
This is of particular importance for the 3rd generation of GW detectors for which
we expect to see postmerger signals with SNRs of $\sim 10$. 
Realistic sources assuming an advanced LIGO sensitivity will 
only have SNRs of about 2-3, e.g.~\cite{Dudi:2018jzn,Tsang:2018uie}.

In Fig.~\ref{fig:GW_spectrum}, we show the postmerger spectrogram of our simulations 
plotted with respect to the retarded time, e.g., Eq.~(22) of~\cite{Dietrich:2015iva}. 
The spectrograms include individual modes up to $(\ell,m)=(4,4)$ and we assume an inclination 
of $\iota=\pi/4$ (other inclinations have been tested and resulted in similar results and conclusions). 
Setups which incorporate cooling via nucleon-nucleon-axion bremsstrahlung are shown as colored contours. 
Setups without axion cooling are shown as black contour lines. 
The plot shows a comparison between the most extreme cases with $m_{a}=1\rm eV$
and $m_{a}=0\rm eV$. We find that the differences observed in Fig.~\ref{fig:HM4235} lead 
to very small difference in the postmerger GW spectrogram. 
Overall the shift in the main postmerger GW emission frequency is significantly below $50\rm Hz$, 
thus, we do not expect that available techniques as discussed in 
Refs.~\cite{Chatziioannou:2017ixj,Tsang:2018uie} will be able to place constraints on the axion mass even 
for a hypothetical mass of $m_{a}=1\rm eV$. As noted above, at higher masses 
a correct treatment of axion reabsorption would 
reduce the cooling effect, meaning that our $m_{a}=1\rm eV$ simulation is already a ``best case'' scenario.

Thus, we conclude that the impact of
nucleon-nucleon-axion bremsstrahlung on the
postmerger GW signal will not provide a new way of constraining the axion mass
beyond the current existing mass limits. 

\subsection{Electromagnetic emission}

\begin{figure}[t]
 \centering
 \includegraphics[width=\columnwidth]{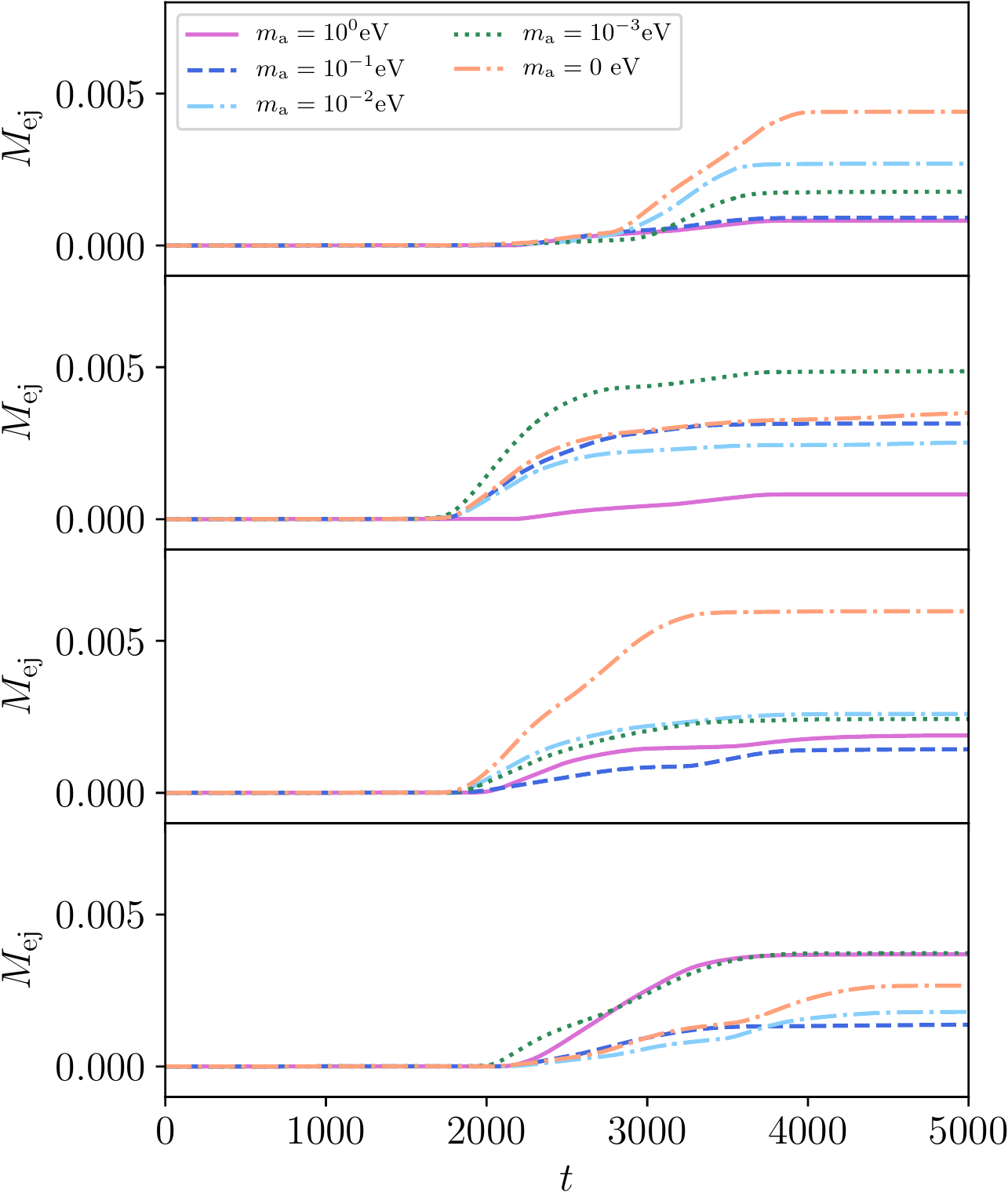}
 \caption{Mass of the ejected material for all our simulations with various total masses (different panels) 
 and axion masses (different line colors and styles).}
 \label{fig:ejecta}
\end{figure}

Given the multi-messenger strategy of observing BNS mergers, we also consider the possibility that the
nucleon-nucleon-axion bremsstrahlung cooling affects electromagnetic signatures related to BNS mergers, 
where in particular we will focus on the kilonova detectable in the optical, infrared, 
and ultraviolet, see \cite{Metzger:2016pju,Tanaka:2016sbx} for reviews and further references. 
The kilonovae are triggered by the radioactive decay of r-process nuclei 
in the neutron-rich material ejected during a BNS merger, thus, we will investigate 
in more detail the ejecta mechanism and the amount of ejecta during and after the collision of the two 
neutron.
We show in Fig.~\ref{fig:ejecta} for all our simulations the total mass of the ejected material. 
The ejecta mass is estimated as the unbound material leaving an extraction sphere 
of a radius of $r_{\rm ej} = 200$. 
We find that for most configurations with small axion cooling effects, i.e., small axion masses, 
the dynamical ejecta is larger than for setups for which an additional cooling via axions is incorporated. 
However, this behavior is not monotonic and more based on a qualitative than a quantitative inspection of the results. 
An interesting and robust feature in all our simulations with an axion mass of $m_{a}=1\rm eV$ is 
that with an increasing total mass, the mass of the ejected material decreases. 
This is an indicator of the the fact that for large axion masses, tidal ejecta 
are the dominant ejection mechanism. Shock driven ejecta are less important, which 
is in agreement with the naive expectation that the additional axion cooling reduces 
the thermal pressure. 
Therefore, since an increasing total mass of the binary leads to a smaller tidal deformabilities,  
an ejection due to tidal effects (torque in the tidal tail) is less likely. 

In both numerical simulations and analyses of the measured kilonova lightcurves, 
the uncertainty of the ejecta mass is larger than the small 
differences found in our work (see e.g.~\cite{Dietrich:2015iva,Shibata:2017xdx,Radice:2018pdn} 
for estimate about ejecta uncertainties).  
We conclude that for our configurations the incorporation of axion cooling will 
not lead to a measurable effect, even for a hypothetical axion mass of $m_{a}=1\rm eV$. 

\section{Summary and Conclusion}
\label{sec:summary}

In this article, we have presented the first simulations which incorporate 
in a covariant and phenomenological way nucleon-nucleon-axion bremsstrahlung effects. 
These provide an additional cooling channel for the postmerger remnant and therefore 
could have an influence on the postmerger dynamics. 
We find interesting differences in our simulations, 
in particular, that with an increasing mass of the axion, 
the lifetime of the remnant before collapsing to a BH decreases, 
which is caused by the missing temperature support, and that
a faster sphericalisation of the remnant occurs. 
However, this shift in the lifetime is of the order of a few milliseconds and, while it might be more pronounced for 
configurations with a longer remnant lifetime, with current technology for observing BNS events
it would not lead to a noticeable effect. 
We also studied the emitted postmerger GW signal and the amount of ejected material 
during and after the collision of the two stars. 
We find noticeable differences between simulations
with and without axion cooling, but the change is insufficient to
be detectable from GW and EM observations. 
Even for the upcoming 3rd generation of 
GW detectors, it seems unlikely that the axion mass could be constrained 
via the effect of nucleon-nucleon-axion bremsstrahlung. 

We note that we restrict our work to a single hadronic equation of state and to 
four different equal-mass configurations. We also neglect neutrinos, and assume a simplified cooling model, without
taking into account reabsorption, which should play a role in higher mass cases.
It is therefore possible that other equations of state, more accurate models, 
or different configuration setups might lead to larger, more noticeable effects. 
However, our work indicates that constraints on the axion mass cannot easily be derived 
from the additional axion cooling channel provided during the postmerger.
We note that there are other effects which might constrain the mass of the axion or ALPs
with the help 
of multi-messenger detections of BNS mergers. For example, a change in the inspiral GW signal, 
e.g.,~\cite{Brito:2017zvb, Blas:2016ddr, Rozner:2019gba, Hook:2017psm,Huang:2018pbu} or 
axion-photon conversion in the presence of magnetic fields, 
e.g.,~\cite{Raffelt:1987np,Iwazaki:2014wka,Bai:2017feq,Dietrich:2018jov}. 

\begin{acknowledgments}
TD thanks Pedro Ferreira and the members the Beecroft Institute at the 
University Oxford for hospitality while working on this project. 
TD acknowledges support by the European Union's Horizon 2020 research and 
innovation program under grant agreement No 749145, BNSmergers.
KC acknowledges support from the European Research Council 
under grant agreement ID 693024, GravityLS.
Computations have been performed on the Minerva cluster of 
the Max-Planck Institute for Gravitational Physics.
\end{acknowledgments}

\appendix 

\begin{figure}[t]
 \centering
 \includegraphics[width=\columnwidth]{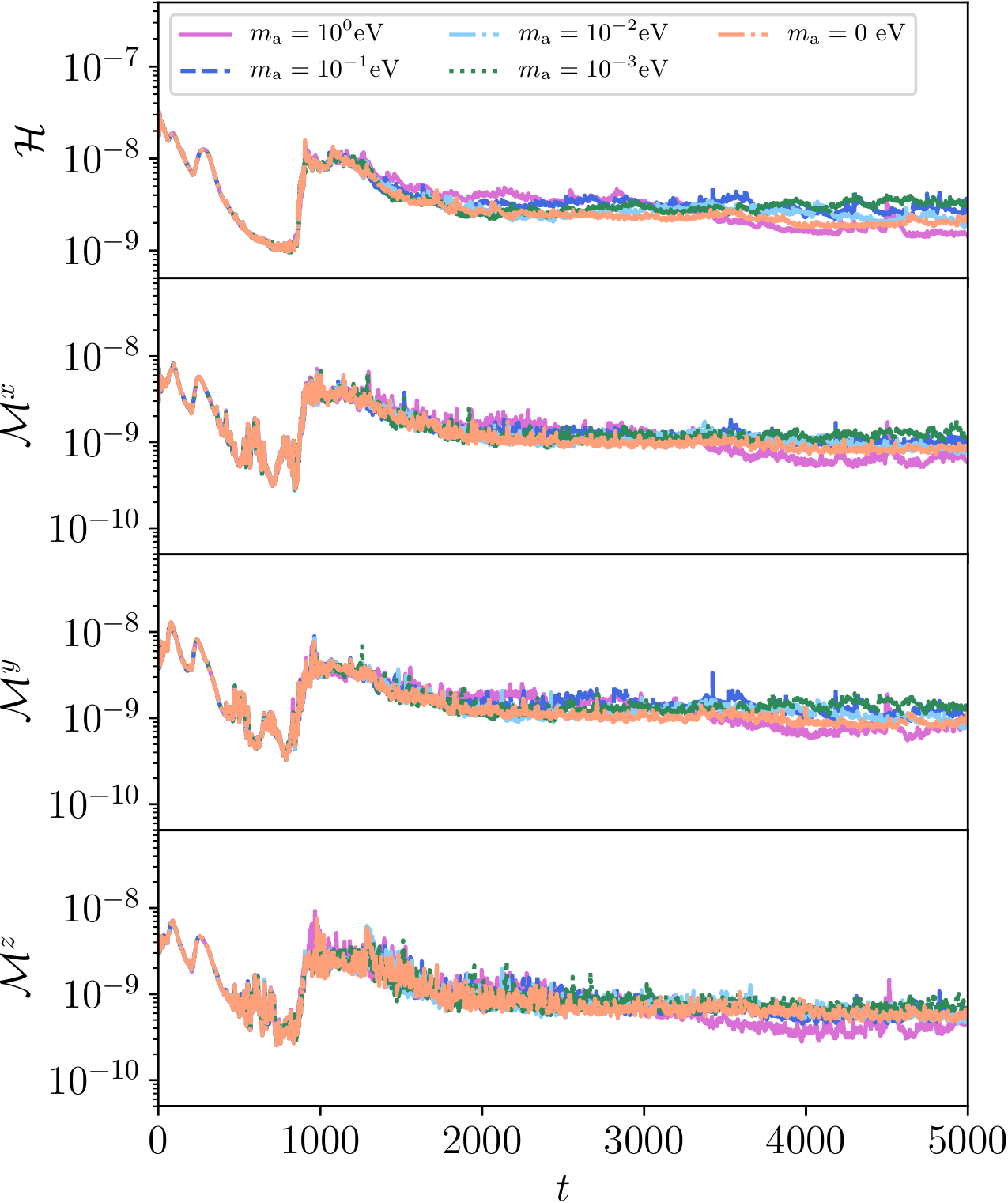}
 \caption{Hamiltonian (top panel) and Momentum Constraint violations (second to forth panel) 
 for different axion masses.  }
 \label{fig:constraints}
\end{figure}

\section{Unit conversion}
\label{app:units}
Based on our own experience, we find it useful to shortly present the necessary unit conversions 
needed to transform expressions from natural or cgs units to geometric units 
as used in our simulations ($M_\odot=c=G=1$):
\begin{eqnarray}
 1\rm cm & \rightarrow & \frac{1}{1.476625 \times 10^5} M_\odot = 6.772\times 10^{-6},\\ 
 1 \rm g & \rightarrow & \frac{1}{1.98847\times10^{33}} M_\odot = 5.029 \times 10^{-34},\\
 1 \rm s & \rightarrow & \frac{1}{4.92569\times 10^{-6}} M_\odot = 2.030 \times 10^5, 
\end{eqnarray}
furthermore: $\rm [erg cm^{-3} s^{-1}] = g^1 s^{-3} cm^{-1}$.
Note further that from $1\rm eV = 11605\rm K$. 
Our particular choice of $M_\odot=c=G=1$, as employed in BAM, 
also determines all units in Figs.~\ref{fig:HM4235}-\ref{fig:constraints}.

\section{Constraint violations}
\label{app:constraints}
To ensure the consistency of our simulations and that the phenomenological 
description of the nucleon-nucleon-axion bremsstrahlung is in agreement with general relativity, 
we will present the time evolution of the Einstein constraints. 
Figure~\ref{fig:constraints} shows the evolution of the Hamiltonian Constraint $\mathcal{H}$ and 
the individual components of the Momentum Constraint $\mathcal{M}^i$. 
As an example, we show the evolution of the Case-2 configuration 
and evaluate the constraints on the refinement level $l=2$. 
This level is non-moving, which allows a one-to-one comparison between all simulations, 
and extends from $[-293,293]\times[-293,293]\times[-293,-15]$ in $x\times y \times z$-directions; 
note the employed bitant symmetry in z-direction. 
Other configurations show a similar behavior. 

In all simulations, we find no indication that the additional axion-cooling channel 
leads to an increase of the constraint violations. 
This observation validates the correctness of our approach.

\bibliography{paper20190903.bbl}

\begin{thebibliography}{91}
\expandafter\ifx\csname natexlab\endcsname\relax\def\natexlab#1{#1}\fi
\expandafter\ifx\csname bibnamefont\endcsname\relax
  \def\bibnamefont#1{#1}\fi
\expandafter\ifx\csname bibfnamefont\endcsname\relax
  \def\bibfnamefont#1{#1}\fi
\expandafter\ifx\csname citenamefont\endcsname\relax
  \def\citenamefont#1{#1}\fi
\expandafter\ifx\csname url\endcsname\relax
  \def\url#1{\texttt{#1}}\fi
\expandafter\ifx\csname urlprefix\endcsname\relax\def\urlprefix{URL }\fi
\providecommand{\bibinfo}[2]{#2}
\providecommand{\eprint}[2][]{\url{#2}}

\bibitem[{\citenamefont{Abbott
  et~al.}(2017{\natexlab{a}})}]{TheLIGOScientific:2017qsa}
\bibinfo{author}{\bibfnamefont{B.~P.} \bibnamefont{Abbott}}
  \bibnamefont{et~al.} (\bibinfo{collaboration}{Virgo, LIGO Scientific}),
  \bibinfo{journal}{Phys. Rev. Lett.} \textbf{\bibinfo{volume}{119}},
  \bibinfo{pages}{161101} (\bibinfo{year}{2017}{\natexlab{a}}),
  \eprint{1710.05832}.

\bibitem[{\citenamefont{Abbott et~al.}(2018{\natexlab{a}})}]{Abbott:2018exr}
\bibinfo{author}{\bibfnamefont{B.~P.} \bibnamefont{Abbott}}
  \bibnamefont{et~al.} (\bibinfo{collaboration}{Virgo, LIGO Scientific}),
  \bibinfo{journal}{Phys. Rev. Lett.} \textbf{\bibinfo{volume}{121}},
  \bibinfo{pages}{161101} (\bibinfo{year}{2018}{\natexlab{a}}),
  \eprint{1805.11581}.

\bibitem[{\citenamefont{Abbott et~al.}(2019)}]{Abbott:2018wiz}
\bibinfo{author}{\bibfnamefont{B.~P.} \bibnamefont{Abbott}}
  \bibnamefont{et~al.} (\bibinfo{collaboration}{LIGO Scientific, Virgo}),
  \bibinfo{journal}{Phys. Rev.} \textbf{\bibinfo{volume}{X9}},
  \bibinfo{pages}{011001} (\bibinfo{year}{2019}), \eprint{1805.11579}.

\bibitem[{\citenamefont{Abbott
  et~al.}(2018{\natexlab{b}})}]{LIGOScientific:2018mvr}
\bibinfo{author}{\bibfnamefont{B.~P.} \bibnamefont{Abbott}}
  \bibnamefont{et~al.} (\bibinfo{collaboration}{LIGO Scientific, Virgo})
  (\bibinfo{year}{2018}{\natexlab{b}}), \eprint{1811.12907}.

\bibitem[{\citenamefont{Abbott et~al.}(2017{\natexlab{b}})}]{Monitor:2017mdv}
\bibinfo{author}{\bibfnamefont{B.~P.} \bibnamefont{Abbott}}
  \bibnamefont{et~al.} (\bibinfo{collaboration}{Virgo, Fermi-GBM, INTEGRAL,
  LIGO Scientific}), \bibinfo{journal}{Astrophys. J.}
  \textbf{\bibinfo{volume}{848}}, \bibinfo{pages}{L13}
  (\bibinfo{year}{2017}{\natexlab{b}}), \eprint{1710.05834}.

\bibitem[{GBM(2017)}]{GBM:2017lvd}
\bibinfo{journal}{Astrophys. J.} \textbf{\bibinfo{volume}{848}},
  \bibinfo{pages}{L12} (\bibinfo{year}{2017}), \eprint{1710.05833}.

\bibitem[{\citenamefont{Coulter et~al.}(2017)}]{Coulter:2017wya}
\bibinfo{author}{\bibfnamefont{D.~A.} \bibnamefont{Coulter}}
  \bibnamefont{et~al.}, \bibinfo{journal}{Science}  (\bibinfo{year}{2017}),
  \bibinfo{note}{[Science358,1556(2017)]}, \eprint{1710.05452}.

\bibitem[{\citenamefont{Abbott et~al.}(2016)}]{Abbott:2016ymx}
\bibinfo{author}{\bibfnamefont{B.~P.} \bibnamefont{Abbott}}
  \bibnamefont{et~al.} (\bibinfo{collaboration}{Virgo, LIGO Scientific}),
  \bibinfo{journal}{Astrophys. J.} \textbf{\bibinfo{volume}{832}},
  \bibinfo{pages}{L21} (\bibinfo{year}{2016}), \eprint{1607.07456}.

\bibitem[{\citenamefont{Chruslinska et~al.}(2018)\citenamefont{Chruslinska,
  Belczynski, Klencki, and Benacquista}}]{Chruslinska:2017odi}
\bibinfo{author}{\bibfnamefont{M.}~\bibnamefont{Chruslinska}},
  \bibinfo{author}{\bibfnamefont{K.}~\bibnamefont{Belczynski}},
  \bibinfo{author}{\bibfnamefont{J.}~\bibnamefont{Klencki}}, \bibnamefont{and}
  \bibinfo{author}{\bibfnamefont{M.}~\bibnamefont{Benacquista}},
  \bibinfo{journal}{Mon. Not. Roy. Astron. Soc.}
  \textbf{\bibinfo{volume}{474}}, \bibinfo{pages}{2937} (\bibinfo{year}{2018}),
  \eprint{1708.07885}.

\bibitem[{\citenamefont{Peccei and Quinn}(1977)}]{Peccei:1977hh}
\bibinfo{author}{\bibfnamefont{R.~D.} \bibnamefont{Peccei}} \bibnamefont{and}
  \bibinfo{author}{\bibfnamefont{H.~R.} \bibnamefont{Quinn}},
  \bibinfo{journal}{Phys. Rev. Lett.} \textbf{\bibinfo{volume}{38}},
  \bibinfo{pages}{1440} (\bibinfo{year}{1977}), \bibinfo{note}{[,328(1977)]}.

\bibitem[{\citenamefont{Vafa and Witten}(1984)}]{Vafa:1984xg}
\bibinfo{author}{\bibfnamefont{C.}~\bibnamefont{Vafa}} \bibnamefont{and}
  \bibinfo{author}{\bibfnamefont{E.}~\bibnamefont{Witten}},
  \bibinfo{journal}{Phys. Rev. Lett.} \textbf{\bibinfo{volume}{53}},
  \bibinfo{pages}{535} (\bibinfo{year}{1984}).

\bibitem[{\citenamefont{Hook}(2019)}]{Hook:2018dlk}
\bibinfo{author}{\bibfnamefont{A.}~\bibnamefont{Hook}}, \bibinfo{journal}{PoS}
  \textbf{\bibinfo{volume}{TASI2018}}, \bibinfo{pages}{004}
  (\bibinfo{year}{2019}), \eprint{1812.02669}.

\bibitem[{\citenamefont{Svrcek and Witten}(2006)}]{Svrcek:2006yi}
\bibinfo{author}{\bibfnamefont{P.}~\bibnamefont{Svrcek}} \bibnamefont{and}
  \bibinfo{author}{\bibfnamefont{E.}~\bibnamefont{Witten}},
  \bibinfo{journal}{JHEP} \textbf{\bibinfo{volume}{06}}, \bibinfo{pages}{051}
  (\bibinfo{year}{2006}), \eprint{hep-th/0605206}.

\bibitem[{\citenamefont{Kim}(1987)}]{Kim:1986ax}
\bibinfo{author}{\bibfnamefont{J.~E.} \bibnamefont{Kim}},
  \bibinfo{journal}{Phys. Rept.} \textbf{\bibinfo{volume}{150}},
  \bibinfo{pages}{1} (\bibinfo{year}{1987}).

\bibitem[{\citenamefont{Preskill et~al.}(1983)\citenamefont{Preskill, Wise, and
  Wilczek}}]{Preskill:1982cy}
\bibinfo{author}{\bibfnamefont{J.}~\bibnamefont{Preskill}},
  \bibinfo{author}{\bibfnamefont{M.~B.} \bibnamefont{Wise}}, \bibnamefont{and}
  \bibinfo{author}{\bibfnamefont{F.}~\bibnamefont{Wilczek}},
  \bibinfo{journal}{Phys. Lett.} \textbf{\bibinfo{volume}{B120}},
  \bibinfo{pages}{127} (\bibinfo{year}{1983}), \bibinfo{note}{[,URL(1982)]}.

\bibitem[{\citenamefont{Abbott and Sikivie}(1983)}]{Abbott:1982af}
\bibinfo{author}{\bibfnamefont{L.~F.} \bibnamefont{Abbott}} \bibnamefont{and}
  \bibinfo{author}{\bibfnamefont{P.}~\bibnamefont{Sikivie}},
  \bibinfo{journal}{Phys. Lett.} \textbf{\bibinfo{volume}{B120}},
  \bibinfo{pages}{133} (\bibinfo{year}{1983}), \bibinfo{note}{[,URL(1982)]}.

\bibitem[{\citenamefont{Dine and Fischler}(1983)}]{Dine:1982ah}
\bibinfo{author}{\bibfnamefont{M.}~\bibnamefont{Dine}} \bibnamefont{and}
  \bibinfo{author}{\bibfnamefont{W.}~\bibnamefont{Fischler}},
  \bibinfo{journal}{Phys. Lett.} \textbf{\bibinfo{volume}{B120}},
  \bibinfo{pages}{137} (\bibinfo{year}{1983}), \bibinfo{note}{[,URL(1982)]}.

\bibitem[{\citenamefont{Marsh}(2016)}]{Marsh:2015xka}
\bibinfo{author}{\bibfnamefont{D.~J.~E.} \bibnamefont{Marsh}},
  \bibinfo{journal}{Phys. Rept.} \textbf{\bibinfo{volume}{643}},
  \bibinfo{pages}{1} (\bibinfo{year}{2016}), \eprint{1510.07633}.

\bibitem[{\citenamefont{Brinkmann and Turner}(1988)}]{Brinkmann:1988vi}
\bibinfo{author}{\bibfnamefont{R.~P.} \bibnamefont{Brinkmann}}
  \bibnamefont{and} \bibinfo{author}{\bibfnamefont{M.~S.}
  \bibnamefont{Turner}}, \bibinfo{journal}{Phys. Rev.}
  \textbf{\bibinfo{volume}{D38}}, \bibinfo{pages}{2338} (\bibinfo{year}{1988}).

\bibitem[{\citenamefont{Dicus et~al.}(1978)\citenamefont{Dicus, Kolb, Teplitz,
  and Wagoner}}]{Dicus:1978fp}
\bibinfo{author}{\bibfnamefont{D.~A.} \bibnamefont{Dicus}},
  \bibinfo{author}{\bibfnamefont{E.~W.} \bibnamefont{Kolb}},
  \bibinfo{author}{\bibfnamefont{V.~L.} \bibnamefont{Teplitz}},
  \bibnamefont{and} \bibinfo{author}{\bibfnamefont{R.~V.}
  \bibnamefont{Wagoner}}, \bibinfo{journal}{Phys. Rev.}
  \textbf{\bibinfo{volume}{D18}}, \bibinfo{pages}{1829} (\bibinfo{year}{1978}).

\bibitem[{\citenamefont{Vysotsky et~al.}(1978)\citenamefont{Vysotsky,
  Zeldovich, Khlopov, and Chechetkin}}]{Vysotsky:1978dc}
\bibinfo{author}{\bibfnamefont{M.~I.} \bibnamefont{Vysotsky}},
  \bibinfo{author}{\bibfnamefont{{\relax Ya}.~B.} \bibnamefont{Zeldovich}},
  \bibinfo{author}{\bibfnamefont{M.~{\relax Yu}.} \bibnamefont{Khlopov}},
  \bibnamefont{and} \bibinfo{author}{\bibfnamefont{V.~M.}
  \bibnamefont{Chechetkin}}, \bibinfo{journal}{Pisma Zh. Eksp. Teor. Fiz.}
  \textbf{\bibinfo{volume}{27}}, \bibinfo{pages}{533} (\bibinfo{year}{1978}),
  \bibinfo{note}{[JETP Lett.27,502(1978)]}.

\bibitem[{\citenamefont{Raffelt}(1999)}]{Raffelt:1999tx}
\bibinfo{author}{\bibfnamefont{G.~G.} \bibnamefont{Raffelt}},
  \bibinfo{journal}{Ann. Rev. Nucl. Part. Sci.} \textbf{\bibinfo{volume}{49}},
  \bibinfo{pages}{163} (\bibinfo{year}{1999}), \eprint{hep-ph/9903472}.

\bibitem[{\citenamefont{Raffelt}(1988)}]{Raffelt:1987np}
\bibinfo{author}{\bibfnamefont{G.~G.} \bibnamefont{Raffelt}},
  \bibinfo{journal}{Phys. Rev.} \textbf{\bibinfo{volume}{D37}},
  \bibinfo{pages}{1356} (\bibinfo{year}{1988}).

\bibitem[{\citenamefont{Schlattl et~al.}(1999)\citenamefont{Schlattl, Weiss,
  and Raffelt}}]{Schlattl:1998fz}
\bibinfo{author}{\bibfnamefont{H.}~\bibnamefont{Schlattl}},
  \bibinfo{author}{\bibfnamefont{A.}~\bibnamefont{Weiss}}, \bibnamefont{and}
  \bibinfo{author}{\bibfnamefont{G.}~\bibnamefont{Raffelt}},
  \bibinfo{journal}{Astropart. Phys.} \textbf{\bibinfo{volume}{10}},
  \bibinfo{pages}{353} (\bibinfo{year}{1999}), \eprint{hep-ph/9807476}.

\bibitem[{\citenamefont{Isern and Garcia-Berro}(2003)}]{Isern:2003xj}
\bibinfo{author}{\bibfnamefont{J.}~\bibnamefont{Isern}} \bibnamefont{and}
  \bibinfo{author}{\bibfnamefont{E.}~\bibnamefont{Garcia-Berro}},
  \bibinfo{journal}{Nucl. Phys. Proc. Suppl.} \textbf{\bibinfo{volume}{114}},
  \bibinfo{pages}{107} (\bibinfo{year}{2003}), \bibinfo{note}{[,107(2003)]}.

\bibitem[{\citenamefont{Raffelt}(2008)}]{Raffelt:2006cw}
\bibinfo{author}{\bibfnamefont{G.~G.} \bibnamefont{Raffelt}},
  \bibinfo{journal}{Lect. Notes Phys.} \textbf{\bibinfo{volume}{741}},
  \bibinfo{pages}{51} (\bibinfo{year}{2008}), \bibinfo{note}{[,51(2006)]},
  \eprint{hep-ph/0611350}.

\bibitem[{\citenamefont{Ellis and Olive}(1987)}]{Ellis:1987pk}
\bibinfo{author}{\bibfnamefont{J.~R.} \bibnamefont{Ellis}} \bibnamefont{and}
  \bibinfo{author}{\bibfnamefont{K.~A.} \bibnamefont{Olive}},
  \bibinfo{journal}{Phys. Lett.} \textbf{\bibinfo{volume}{B193}},
  \bibinfo{pages}{525} (\bibinfo{year}{1987}).

\bibitem[{\citenamefont{Raffelt and Seckel}(1988)}]{Raffelt:1987yt}
\bibinfo{author}{\bibfnamefont{G.}~\bibnamefont{Raffelt}} \bibnamefont{and}
  \bibinfo{author}{\bibfnamefont{D.}~\bibnamefont{Seckel}},
  \bibinfo{journal}{Phys. Rev. Lett.} \textbf{\bibinfo{volume}{60}},
  \bibinfo{pages}{1793} (\bibinfo{year}{1988}).

\bibitem[{\citenamefont{Turner}(1988)}]{Turner:1987by}
\bibinfo{author}{\bibfnamefont{M.~S.} \bibnamefont{Turner}},
  \bibinfo{journal}{Phys. Rev. Lett.} \textbf{\bibinfo{volume}{60}},
  \bibinfo{pages}{1797} (\bibinfo{year}{1988}).

\bibitem[{\citenamefont{Mayle et~al.}(1988)\citenamefont{Mayle, Wilson, Ellis,
  Olive, Schramm, and Steigman}}]{Mayle:1987as}
\bibinfo{author}{\bibfnamefont{R.}~\bibnamefont{Mayle}},
  \bibinfo{author}{\bibfnamefont{J.~R.} \bibnamefont{Wilson}},
  \bibinfo{author}{\bibfnamefont{J.~R.} \bibnamefont{Ellis}},
  \bibinfo{author}{\bibfnamefont{K.~A.} \bibnamefont{Olive}},
  \bibinfo{author}{\bibfnamefont{D.~N.} \bibnamefont{Schramm}},
  \bibnamefont{and} \bibinfo{author}{\bibfnamefont{G.}~\bibnamefont{Steigman}},
  \bibinfo{journal}{Phys. Lett.} \textbf{\bibinfo{volume}{B203}},
  \bibinfo{pages}{188} (\bibinfo{year}{1988}).

\bibitem[{\citenamefont{Mayle et~al.}(1989)\citenamefont{Mayle, Wilson, Ellis,
  Olive, Schramm, and Steigman}}]{Mayle:1989yx}
\bibinfo{author}{\bibfnamefont{R.}~\bibnamefont{Mayle}},
  \bibinfo{author}{\bibfnamefont{J.~R.} \bibnamefont{Wilson}},
  \bibinfo{author}{\bibfnamefont{J.~R.} \bibnamefont{Ellis}},
  \bibinfo{author}{\bibfnamefont{K.~A.} \bibnamefont{Olive}},
  \bibinfo{author}{\bibfnamefont{D.~N.} \bibnamefont{Schramm}},
  \bibnamefont{and} \bibinfo{author}{\bibfnamefont{G.}~\bibnamefont{Steigman}},
  \bibinfo{journal}{Phys. Lett.} \textbf{\bibinfo{volume}{B219}},
  \bibinfo{pages}{515} (\bibinfo{year}{1989}), \bibinfo{note}{[,188(1989)]}.

\bibitem[{\citenamefont{Burrows et~al.}(1989)\citenamefont{Burrows, Turner, and
  Brinkmann}}]{Burrows:1988ah}
\bibinfo{author}{\bibfnamefont{A.}~\bibnamefont{Burrows}},
  \bibinfo{author}{\bibfnamefont{M.~S.} \bibnamefont{Turner}},
  \bibnamefont{and} \bibinfo{author}{\bibfnamefont{R.~P.}
  \bibnamefont{Brinkmann}}, \bibinfo{journal}{Phys. Rev.}
  \textbf{\bibinfo{volume}{D39}}, \bibinfo{pages}{1020} (\bibinfo{year}{1989}).

\bibitem[{\citenamefont{Burrows et~al.}(1990)\citenamefont{Burrows, Ressell,
  and Turner}}]{Burrows:1990pk}
\bibinfo{author}{\bibfnamefont{A.}~\bibnamefont{Burrows}},
  \bibinfo{author}{\bibfnamefont{M.~T.} \bibnamefont{Ressell}},
  \bibnamefont{and} \bibinfo{author}{\bibfnamefont{M.~S.}
  \bibnamefont{Turner}}, \bibinfo{journal}{Phys. Rev.}
  \textbf{\bibinfo{volume}{D42}}, \bibinfo{pages}{3297} (\bibinfo{year}{1990}).

\bibitem[{\citenamefont{Keil et~al.}(1997)\citenamefont{Keil, Janka, Schramm,
  Sigl, Turner, and Ellis}}]{Keil:1996ju}
\bibinfo{author}{\bibfnamefont{W.}~\bibnamefont{Keil}},
  \bibinfo{author}{\bibfnamefont{H.-T.} \bibnamefont{Janka}},
  \bibinfo{author}{\bibfnamefont{D.~N.} \bibnamefont{Schramm}},
  \bibinfo{author}{\bibfnamefont{G.}~\bibnamefont{Sigl}},
  \bibinfo{author}{\bibfnamefont{M.~S.} \bibnamefont{Turner}},
  \bibnamefont{and} \bibinfo{author}{\bibfnamefont{J.~R.} \bibnamefont{Ellis}},
  \bibinfo{journal}{Phys. Rev.} \textbf{\bibinfo{volume}{D56}},
  \bibinfo{pages}{2419} (\bibinfo{year}{1997}), \eprint{astro-ph/9612222}.

\bibitem[{\citenamefont{Chang et~al.}(2018)\citenamefont{Chang, Essig, and
  McDermott}}]{Chang:2018rso}
\bibinfo{author}{\bibfnamefont{J.~H.} \bibnamefont{Chang}},
  \bibinfo{author}{\bibfnamefont{R.}~\bibnamefont{Essig}}, \bibnamefont{and}
  \bibinfo{author}{\bibfnamefont{S.~D.} \bibnamefont{McDermott}},
  \bibinfo{journal}{JHEP} \textbf{\bibinfo{volume}{09}}, \bibinfo{pages}{051}
  (\bibinfo{year}{2018}), \eprint{1803.00993}.

\bibitem[{\citenamefont{Paschalidis et~al.}(2012)\citenamefont{Paschalidis,
  Etienne, and Shapiro}}]{Paschalidis:2012ff}
\bibinfo{author}{\bibfnamefont{V.}~\bibnamefont{Paschalidis}},
  \bibinfo{author}{\bibfnamefont{Z.~B.} \bibnamefont{Etienne}},
  \bibnamefont{and} \bibinfo{author}{\bibfnamefont{S.~L.}
  \bibnamefont{Shapiro}}, \bibinfo{journal}{Phys.Rev.}
  \textbf{\bibinfo{volume}{D86}}, \bibinfo{pages}{064032}
  (\bibinfo{year}{2012}), \eprint{1208.5487}.

\bibitem[{\citenamefont{Gill et~al.}(2019)\citenamefont{Gill, Nathanail, and
  Rezzolla}}]{Gill:2019bvq}
\bibinfo{author}{\bibfnamefont{R.}~\bibnamefont{Gill}},
  \bibinfo{author}{\bibfnamefont{A.}~\bibnamefont{Nathanail}},
  \bibnamefont{and} \bibinfo{author}{\bibfnamefont{L.}~\bibnamefont{Rezzolla}},
  \bibinfo{journal}{Astrophys. J.} \textbf{\bibinfo{volume}{876}},
  \bibinfo{pages}{139} (\bibinfo{year}{2019}), \eprint{1901.04138}.

\bibitem[{\citenamefont{Abbott et~al.}(2017{\natexlab{c}})}]{Abbott:2017dke}
\bibinfo{author}{\bibfnamefont{B.~P.} \bibnamefont{Abbott}}
  \bibnamefont{et~al.} (\bibinfo{collaboration}{Virgo, LIGO Scientific}),
  \bibinfo{journal}{Astrophys. J.} \textbf{\bibinfo{volume}{851}},
  \bibinfo{pages}{L16} (\bibinfo{year}{2017}{\natexlab{c}}),
  \eprint{1710.09320}.

\bibitem[{\citenamefont{Rosswog and Liebendoerfer}(2003)}]{Rosswog:2003rv}
\bibinfo{author}{\bibfnamefont{S.}~\bibnamefont{Rosswog}} \bibnamefont{and}
  \bibinfo{author}{\bibfnamefont{M.}~\bibnamefont{Liebendoerfer}},
  \bibinfo{journal}{Mon.Not.Roy.Astron.Soc.} \textbf{\bibinfo{volume}{342}},
  \bibinfo{pages}{673} (\bibinfo{year}{2003}), \eprint{astro-ph/0302301}.

\bibitem[{\citenamefont{Dessart et~al.}(2009)\citenamefont{Dessart, Ott,
  Burrows, Rosswog, and Livne}}]{Dessart:2008zd}
\bibinfo{author}{\bibfnamefont{L.}~\bibnamefont{Dessart}},
  \bibinfo{author}{\bibfnamefont{C.}~\bibnamefont{Ott}},
  \bibinfo{author}{\bibfnamefont{A.}~\bibnamefont{Burrows}},
  \bibinfo{author}{\bibfnamefont{S.}~\bibnamefont{Rosswog}}, \bibnamefont{and}
  \bibinfo{author}{\bibfnamefont{E.}~\bibnamefont{Livne}},
  \bibinfo{journal}{Astrophys.J.} \textbf{\bibinfo{volume}{690}},
  \bibinfo{pages}{1681} (\bibinfo{year}{2009}), \eprint{0806.4380}.

\bibitem[{\citenamefont{Sekiguchi et~al.}(2011)\citenamefont{Sekiguchi, Kiuchi,
  Kyutoku, and Shibata}}]{Sekiguchi:2011zd}
\bibinfo{author}{\bibfnamefont{Y.}~\bibnamefont{Sekiguchi}},
  \bibinfo{author}{\bibfnamefont{K.}~\bibnamefont{Kiuchi}},
  \bibinfo{author}{\bibfnamefont{K.}~\bibnamefont{Kyutoku}}, \bibnamefont{and}
  \bibinfo{author}{\bibfnamefont{M.}~\bibnamefont{Shibata}},
  \bibinfo{journal}{Phys.Rev.Lett.} \textbf{\bibinfo{volume}{107}},
  \bibinfo{pages}{051102} (\bibinfo{year}{2011}), \eprint{1105.2125}.

\bibitem[{\citenamefont{Galeazzi et~al.}(2013)\citenamefont{Galeazzi, Kastaun,
  Rezzolla, and Font}}]{Galeazzi:2013mia}
\bibinfo{author}{\bibfnamefont{F.}~\bibnamefont{Galeazzi}},
  \bibinfo{author}{\bibfnamefont{W.}~\bibnamefont{Kastaun}},
  \bibinfo{author}{\bibfnamefont{L.}~\bibnamefont{Rezzolla}}, \bibnamefont{and}
  \bibinfo{author}{\bibfnamefont{J.~A.} \bibnamefont{Font}},
  \bibinfo{journal}{Phys.Rev.} \textbf{\bibinfo{volume}{D88}},
  \bibinfo{pages}{064009} (\bibinfo{year}{2013}), \eprint{1306.4953}.

\bibitem[{\citenamefont{Foucart et~al.}(2014)\citenamefont{Foucart, Deaton,
  Duez, O'Connor, Ott, Haas, Kidder, Pfeiffer, Scheel, and
  Szilagyi}}]{Foucart:2014nda}
\bibinfo{author}{\bibfnamefont{F.}~\bibnamefont{Foucart}},
  \bibinfo{author}{\bibfnamefont{M.~B.} \bibnamefont{Deaton}},
  \bibinfo{author}{\bibfnamefont{M.~D.} \bibnamefont{Duez}},
  \bibinfo{author}{\bibfnamefont{E.}~\bibnamefont{O'Connor}},
  \bibinfo{author}{\bibfnamefont{C.~D.} \bibnamefont{Ott}},
  \bibinfo{author}{\bibfnamefont{R.}~\bibnamefont{Haas}},
  \bibinfo{author}{\bibfnamefont{L.~E.} \bibnamefont{Kidder}},
  \bibinfo{author}{\bibfnamefont{H.~P.} \bibnamefont{Pfeiffer}},
  \bibinfo{author}{\bibfnamefont{M.~A.} \bibnamefont{Scheel}},
  \bibnamefont{and} \bibinfo{author}{\bibfnamefont{B.}~\bibnamefont{Szilagyi}},
  \bibinfo{journal}{Phys. Rev.} \textbf{\bibinfo{volume}{D90}},
  \bibinfo{pages}{024026} (\bibinfo{year}{2014}), \eprint{1405.1121}.

\bibitem[{\citenamefont{Neilsen et~al.}(2014)\citenamefont{Neilsen, Liebling,
  Anderson, Lehner, O’Connor et~al.}}]{Neilsen:2014hha}
\bibinfo{author}{\bibfnamefont{D.}~\bibnamefont{Neilsen}},
  \bibinfo{author}{\bibfnamefont{S.~L.} \bibnamefont{Liebling}},
  \bibinfo{author}{\bibfnamefont{M.}~\bibnamefont{Anderson}},
  \bibinfo{author}{\bibfnamefont{L.}~\bibnamefont{Lehner}},
  \bibinfo{author}{\bibfnamefont{E.}~\bibnamefont{O’Connor}},
  \bibnamefont{et~al.}, \bibinfo{journal}{Phys.Rev.}
  \textbf{\bibinfo{volume}{D89}}, \bibinfo{pages}{104029}
  (\bibinfo{year}{2014}), \eprint{1403.3680}.

\bibitem[{\citenamefont{Palenzuela et~al.}(2015)\citenamefont{Palenzuela,
  Liebling, Neilsen, Lehner, Caballero, O’Connor, and
  Anderson}}]{Palenzuela:2015dqa}
\bibinfo{author}{\bibfnamefont{C.}~\bibnamefont{Palenzuela}},
  \bibinfo{author}{\bibfnamefont{S.~L.} \bibnamefont{Liebling}},
  \bibinfo{author}{\bibfnamefont{D.}~\bibnamefont{Neilsen}},
  \bibinfo{author}{\bibfnamefont{L.}~\bibnamefont{Lehner}},
  \bibinfo{author}{\bibfnamefont{O.~L.} \bibnamefont{Caballero}},
  \bibinfo{author}{\bibfnamefont{E.}~\bibnamefont{O’Connor}},
  \bibnamefont{and} \bibinfo{author}{\bibfnamefont{M.}~\bibnamefont{Anderson}},
  \bibinfo{journal}{Phys. Rev.} \textbf{\bibinfo{volume}{D92}},
  \bibinfo{pages}{044045} (\bibinfo{year}{2015}), \eprint{1505.01607}.

\bibitem[{\citenamefont{Lehner et~al.}(2016)\citenamefont{Lehner, Liebling,
  Palenzuela, Caballero, O'Connor, Anderson, and Neilsen}}]{Lehner:2016lxy}
\bibinfo{author}{\bibfnamefont{L.}~\bibnamefont{Lehner}},
  \bibinfo{author}{\bibfnamefont{S.~L.} \bibnamefont{Liebling}},
  \bibinfo{author}{\bibfnamefont{C.}~\bibnamefont{Palenzuela}},
  \bibinfo{author}{\bibfnamefont{O.~L.} \bibnamefont{Caballero}},
  \bibinfo{author}{\bibfnamefont{E.}~\bibnamefont{O'Connor}},
  \bibinfo{author}{\bibfnamefont{M.}~\bibnamefont{Anderson}}, \bibnamefont{and}
  \bibinfo{author}{\bibfnamefont{D.}~\bibnamefont{Neilsen}},
  \bibinfo{journal}{Class. Quant. Grav.} \textbf{\bibinfo{volume}{33}},
  \bibinfo{pages}{184002} (\bibinfo{year}{2016}), \eprint{1603.00501}.

\bibitem[{\citenamefont{Foucart et~al.}(2018)\citenamefont{Foucart, Duez,
  Kidder, Nguyen, Pfeiffer, and Scheel}}]{Foucart:2018gis}
\bibinfo{author}{\bibfnamefont{F.}~\bibnamefont{Foucart}},
  \bibinfo{author}{\bibfnamefont{M.~D.} \bibnamefont{Duez}},
  \bibinfo{author}{\bibfnamefont{L.~E.} \bibnamefont{Kidder}},
  \bibinfo{author}{\bibfnamefont{R.}~\bibnamefont{Nguyen}},
  \bibinfo{author}{\bibfnamefont{H.~P.} \bibnamefont{Pfeiffer}},
  \bibnamefont{and} \bibinfo{author}{\bibfnamefont{M.~A.}
  \bibnamefont{Scheel}}, \bibinfo{journal}{Phys. Rev.}
  \textbf{\bibinfo{volume}{D98}}, \bibinfo{pages}{063007}
  (\bibinfo{year}{2018}), \eprint{1806.02349}.

\bibitem[{\citenamefont{Brito et~al.}(2017)\citenamefont{Brito, Ghosh,
  Barausse, Berti, Cardoso, Dvorkin, Klein, and Pani}}]{Brito:2017zvb}
\bibinfo{author}{\bibfnamefont{R.}~\bibnamefont{Brito}},
  \bibinfo{author}{\bibfnamefont{S.}~\bibnamefont{Ghosh}},
  \bibinfo{author}{\bibfnamefont{E.}~\bibnamefont{Barausse}},
  \bibinfo{author}{\bibfnamefont{E.}~\bibnamefont{Berti}},
  \bibinfo{author}{\bibfnamefont{V.}~\bibnamefont{Cardoso}},
  \bibinfo{author}{\bibfnamefont{I.}~\bibnamefont{Dvorkin}},
  \bibinfo{author}{\bibfnamefont{A.}~\bibnamefont{Klein}}, \bibnamefont{and}
  \bibinfo{author}{\bibfnamefont{P.}~\bibnamefont{Pani}},
  \bibinfo{journal}{Phys. Rev.} \textbf{\bibinfo{volume}{D96}},
  \bibinfo{pages}{064050} (\bibinfo{year}{2017}), \eprint{1706.06311}.

\bibitem[{\citenamefont{Blas et~al.}(2017)\citenamefont{Blas, Nacir, and
  Sibiryakov}}]{Blas:2016ddr}
\bibinfo{author}{\bibfnamefont{D.}~\bibnamefont{Blas}},
  \bibinfo{author}{\bibfnamefont{D.~L.} \bibnamefont{Nacir}}, \bibnamefont{and}
  \bibinfo{author}{\bibfnamefont{S.}~\bibnamefont{Sibiryakov}},
  \bibinfo{journal}{Phys. Rev. Lett.} \textbf{\bibinfo{volume}{118}},
  \bibinfo{pages}{261102} (\bibinfo{year}{2017}), \eprint{1612.06789}.

\bibitem[{\citenamefont{Rozner et~al.}(2019)\citenamefont{Rozner, Grishin,
  Ginat, Igoshev, and Desjacques}}]{Rozner:2019gba}
\bibinfo{author}{\bibfnamefont{M.}~\bibnamefont{Rozner}},
  \bibinfo{author}{\bibfnamefont{E.}~\bibnamefont{Grishin}},
  \bibinfo{author}{\bibfnamefont{Y.~B.} \bibnamefont{Ginat}},
  \bibinfo{author}{\bibfnamefont{A.~P.} \bibnamefont{Igoshev}},
  \bibnamefont{and}
  \bibinfo{author}{\bibfnamefont{V.}~\bibnamefont{Desjacques}}
  (\bibinfo{year}{2019}), \eprint{1904.01958}.

\bibitem[{\citenamefont{Hook and Huang}(2018)}]{Hook:2017psm}
\bibinfo{author}{\bibfnamefont{A.}~\bibnamefont{Hook}} \bibnamefont{and}
  \bibinfo{author}{\bibfnamefont{J.}~\bibnamefont{Huang}},
  \bibinfo{journal}{JHEP} \textbf{\bibinfo{volume}{06}}, \bibinfo{pages}{036}
  (\bibinfo{year}{2018}), \eprint{1708.08464}.

\bibitem[{\citenamefont{Huang et~al.}(2019)\citenamefont{Huang, Johnson,
  Sagunski, Sakellariadou, and Zhang}}]{Huang:2018pbu}
\bibinfo{author}{\bibfnamefont{J.}~\bibnamefont{Huang}},
  \bibinfo{author}{\bibfnamefont{M.~C.} \bibnamefont{Johnson}},
  \bibinfo{author}{\bibfnamefont{L.}~\bibnamefont{Sagunski}},
  \bibinfo{author}{\bibfnamefont{M.}~\bibnamefont{Sakellariadou}},
  \bibnamefont{and} \bibinfo{author}{\bibfnamefont{J.}~\bibnamefont{Zhang}},
  \bibinfo{journal}{Phys. Rev.} \textbf{\bibinfo{volume}{D99}},
  \bibinfo{pages}{063013} (\bibinfo{year}{2019}), \eprint{1807.02133}.

\bibitem[{\citenamefont{Brito et~al.}(2015)\citenamefont{Brito, Cardoso, and
  Pani}}]{Brito:2015oca}
\bibinfo{author}{\bibfnamefont{R.}~\bibnamefont{Brito}},
  \bibinfo{author}{\bibfnamefont{V.}~\bibnamefont{Cardoso}}, \bibnamefont{and}
  \bibinfo{author}{\bibfnamefont{P.}~\bibnamefont{Pani}},
  \bibinfo{journal}{Lect. Notes Phys.} \textbf{\bibinfo{volume}{906}},
  \bibinfo{pages}{pp.1} (\bibinfo{year}{2015}), \eprint{1501.06570}.

\bibitem[{\citenamefont{Brito et~al.}(2016)\citenamefont{Brito, Cardoso,
  Macedo, Okawa, and Palenzuela}}]{Brito:2015yfh}
\bibinfo{author}{\bibfnamefont{R.}~\bibnamefont{Brito}},
  \bibinfo{author}{\bibfnamefont{V.}~\bibnamefont{Cardoso}},
  \bibinfo{author}{\bibfnamefont{C.~F.~B.} \bibnamefont{Macedo}},
  \bibinfo{author}{\bibfnamefont{H.}~\bibnamefont{Okawa}}, \bibnamefont{and}
  \bibinfo{author}{\bibfnamefont{C.}~\bibnamefont{Palenzuela}},
  \bibinfo{journal}{Phys. Rev.} \textbf{\bibinfo{volume}{D93}},
  \bibinfo{pages}{044045} (\bibinfo{year}{2016}), \eprint{1512.00466}.

\bibitem[{\citenamefont{Ellis et~al.}(2018)\citenamefont{Ellis, Hektor, Hütsi,
  Kannike, Marzola, Raidal, and Vaskonen}}]{Ellis:2017jgp}
\bibinfo{author}{\bibfnamefont{J.}~\bibnamefont{Ellis}},
  \bibinfo{author}{\bibfnamefont{A.}~\bibnamefont{Hektor}},
  \bibinfo{author}{\bibfnamefont{G.}~\bibnamefont{Hütsi}},
  \bibinfo{author}{\bibfnamefont{K.}~\bibnamefont{Kannike}},
  \bibinfo{author}{\bibfnamefont{L.}~\bibnamefont{Marzola}},
  \bibinfo{author}{\bibfnamefont{M.}~\bibnamefont{Raidal}}, \bibnamefont{and}
  \bibinfo{author}{\bibfnamefont{V.}~\bibnamefont{Vaskonen}},
  \bibinfo{journal}{Phys. Lett.} \textbf{\bibinfo{volume}{B781}},
  \bibinfo{pages}{607} (\bibinfo{year}{2018}), \eprint{1710.05540}.

\bibitem[{\citenamefont{Bezares et~al.}(2019)\citenamefont{Bezares, Vigano, and
  Palenzuela}}]{Bezares:2019jcb}
\bibinfo{author}{\bibfnamefont{M.}~\bibnamefont{Bezares}},
  \bibinfo{author}{\bibfnamefont{D.}~\bibnamefont{Vigano}}, \bibnamefont{and}
  \bibinfo{author}{\bibfnamefont{C.}~\bibnamefont{Palenzuela}}
  (\bibinfo{year}{2019}), \eprint{1905.08551}.

\bibitem[{\citenamefont{Paschalidis et~al.}(2011)\citenamefont{Paschalidis,
  Liu, Etienne, and Shapiro}}]{Paschalidis:2011ez}
\bibinfo{author}{\bibfnamefont{V.}~\bibnamefont{Paschalidis}},
  \bibinfo{author}{\bibfnamefont{Y.~T.} \bibnamefont{Liu}},
  \bibinfo{author}{\bibfnamefont{Z.}~\bibnamefont{Etienne}}, \bibnamefont{and}
  \bibinfo{author}{\bibfnamefont{S.~L.} \bibnamefont{Shapiro}},
  \bibinfo{journal}{Phys. Rev.} \textbf{\bibinfo{volume}{D84}},
  \bibinfo{pages}{104032} (\bibinfo{year}{2011}), \eprint{1109.5177}.

\bibitem[{\citenamefont{Douchin and Haensel}(2001)}]{Douchin:2001sv}
\bibinfo{author}{\bibfnamefont{F.}~\bibnamefont{Douchin}} \bibnamefont{and}
  \bibinfo{author}{\bibfnamefont{P.}~\bibnamefont{Haensel}},
  \bibinfo{journal}{Astron. Astrophys.} \textbf{\bibinfo{volume}{380}},
  \bibinfo{pages}{151} (\bibinfo{year}{2001}), \eprint{astro-ph/0111092}.

\bibitem[{\citenamefont{Read et~al.}(2009)\citenamefont{Read, Lackey, Owen, and
  Friedman}}]{Read:2008iy}
\bibinfo{author}{\bibfnamefont{J.~S.} \bibnamefont{Read}},
  \bibinfo{author}{\bibfnamefont{B.~D.} \bibnamefont{Lackey}},
  \bibinfo{author}{\bibfnamefont{B.~J.} \bibnamefont{Owen}}, \bibnamefont{and}
  \bibinfo{author}{\bibfnamefont{J.~L.} \bibnamefont{Friedman}},
  \bibinfo{journal}{Phys. Rev.} \textbf{\bibinfo{volume}{D79}},
  \bibinfo{pages}{124032} (\bibinfo{year}{2009}), \eprint{0812.2163}.

\bibitem[{\citenamefont{Radice and Dai}(2019)}]{Radice:2018ozg}
\bibinfo{author}{\bibfnamefont{D.}~\bibnamefont{Radice}} \bibnamefont{and}
  \bibinfo{author}{\bibfnamefont{L.}~\bibnamefont{Dai}}, \bibinfo{journal}{Eur.
  Phys. J.} \textbf{\bibinfo{volume}{A55}}, \bibinfo{pages}{50}
  (\bibinfo{year}{2019}), \eprint{1810.12917}.

\bibitem[{\citenamefont{Coughlin et~al.}(2018)\citenamefont{Coughlin, Dietrich,
  Margalit, and Metzger}}]{Coughlin:2018fis}
\bibinfo{author}{\bibfnamefont{M.~W.} \bibnamefont{Coughlin}},
  \bibinfo{author}{\bibfnamefont{T.}~\bibnamefont{Dietrich}},
  \bibinfo{author}{\bibfnamefont{B.}~\bibnamefont{Margalit}}, \bibnamefont{and}
  \bibinfo{author}{\bibfnamefont{B.~D.} \bibnamefont{Metzger}}
  (\bibinfo{year}{2018}), \eprint{1812.04803}.

\bibitem[{\citenamefont{Tichy}(2009)}]{Tichy:2009yr}
\bibinfo{author}{\bibfnamefont{W.}~\bibnamefont{Tichy}},
  \bibinfo{journal}{Class.Quant.Grav.} \textbf{\bibinfo{volume}{26}},
  \bibinfo{pages}{175018} (\bibinfo{year}{2009}), \eprint{0908.0620}.

\bibitem[{\citenamefont{Tichy}(2012)}]{Tichy:2012rp}
\bibinfo{author}{\bibfnamefont{W.}~\bibnamefont{Tichy}},
  \bibinfo{journal}{Phys. Rev. D} \textbf{\bibinfo{volume}{86}},
  \bibinfo{pages}{064024} (\bibinfo{year}{2012}), \eprint{1209.5336}.

\bibitem[{\citenamefont{Dietrich
  et~al.}(2015{\natexlab{a}})\citenamefont{Dietrich, Moldenhauer,
  Johnson-McDaniel, Bernuzzi, Markakis, Br{\"u}gmann, and
  Tichy}}]{Dietrich:2015pxa}
\bibinfo{author}{\bibfnamefont{T.}~\bibnamefont{Dietrich}},
  \bibinfo{author}{\bibfnamefont{N.}~\bibnamefont{Moldenhauer}},
  \bibinfo{author}{\bibfnamefont{N.~K.} \bibnamefont{Johnson-McDaniel}},
  \bibinfo{author}{\bibfnamefont{S.}~\bibnamefont{Bernuzzi}},
  \bibinfo{author}{\bibfnamefont{C.~M.} \bibnamefont{Markakis}},
  \bibinfo{author}{\bibfnamefont{B.}~\bibnamefont{Br{\"u}gmann}},
  \bibnamefont{and} \bibinfo{author}{\bibfnamefont{W.}~\bibnamefont{Tichy}},
  \bibinfo{journal}{Phys. Rev.} \textbf{\bibinfo{volume}{D92}},
  \bibinfo{pages}{124007} (\bibinfo{year}{2015}{\natexlab{a}}),
  \eprint{1507.07100}.

\bibitem[{\citenamefont{Br{\"u}gmann et~al.}(2008)\citenamefont{Br{\"u}gmann,
  Gonzalez, Hannam, Husa, Sperhake et~al.}}]{Brugmann:2008zz}
\bibinfo{author}{\bibfnamefont{B.}~\bibnamefont{Br{\"u}gmann}},
  \bibinfo{author}{\bibfnamefont{J.~A.} \bibnamefont{Gonzalez}},
  \bibinfo{author}{\bibfnamefont{M.}~\bibnamefont{Hannam}},
  \bibinfo{author}{\bibfnamefont{S.}~\bibnamefont{Husa}},
  \bibinfo{author}{\bibfnamefont{U.}~\bibnamefont{Sperhake}},
  \bibnamefont{et~al.}, \bibinfo{journal}{Phys.Rev.}
  \textbf{\bibinfo{volume}{D77}}, \bibinfo{pages}{024027}
  (\bibinfo{year}{2008}), \eprint{gr-qc/0610128}.

\bibitem[{\citenamefont{Thierfelder et~al.}(2011)\citenamefont{Thierfelder,
  Bernuzzi, and Br{\"u}gmann}}]{Thierfelder:2011yi}
\bibinfo{author}{\bibfnamefont{M.}~\bibnamefont{Thierfelder}},
  \bibinfo{author}{\bibfnamefont{S.}~\bibnamefont{Bernuzzi}}, \bibnamefont{and}
  \bibinfo{author}{\bibfnamefont{B.}~\bibnamefont{Br{\"u}gmann}},
  \bibinfo{journal}{Phys.Rev.} \textbf{\bibinfo{volume}{D84}},
  \bibinfo{pages}{044012} (\bibinfo{year}{2011}), \eprint{1104.4751}.

\bibitem[{\citenamefont{Dietrich
  et~al.}(2015{\natexlab{b}})\citenamefont{Dietrich, Bernuzzi, Ujevic, and
  Br{\"u}gmann}}]{Dietrich:2015iva}
\bibinfo{author}{\bibfnamefont{T.}~\bibnamefont{Dietrich}},
  \bibinfo{author}{\bibfnamefont{S.}~\bibnamefont{Bernuzzi}},
  \bibinfo{author}{\bibfnamefont{M.}~\bibnamefont{Ujevic}}, \bibnamefont{and}
  \bibinfo{author}{\bibfnamefont{B.}~\bibnamefont{Br{\"u}gmann}},
  \bibinfo{journal}{Phys. Rev.} \textbf{\bibinfo{volume}{D91}},
  \bibinfo{pages}{124041} (\bibinfo{year}{2015}{\natexlab{b}}),
  \eprint{1504.01266}.

\bibitem[{\citenamefont{Dietrich et~al.}(2019)\citenamefont{Dietrich, Ossokine,
  and Clough}}]{Dietrich:2018bvi}
\bibinfo{author}{\bibfnamefont{T.}~\bibnamefont{Dietrich}},
  \bibinfo{author}{\bibfnamefont{S.}~\bibnamefont{Ossokine}}, \bibnamefont{and}
  \bibinfo{author}{\bibfnamefont{K.}~\bibnamefont{Clough}},
  \bibinfo{journal}{Class. Quant. Grav.} \textbf{\bibinfo{volume}{36}},
  \bibinfo{pages}{025002} (\bibinfo{year}{2019}), \eprint{1807.06959}.

\bibitem[{\citenamefont{Wilson and Mathews}(1995)}]{Wilson:1995uh}
\bibinfo{author}{\bibfnamefont{J.}~\bibnamefont{Wilson}} \bibnamefont{and}
  \bibinfo{author}{\bibfnamefont{G.}~\bibnamefont{Mathews}},
  \bibinfo{journal}{Phys.Rev.Lett.} \textbf{\bibinfo{volume}{75}},
  \bibinfo{pages}{4161} (\bibinfo{year}{1995}).

\bibitem[{\citenamefont{Wilson et~al.}(1996)\citenamefont{Wilson, Mathews, and
  Marronetti}}]{Wilson:1996ty}
\bibinfo{author}{\bibfnamefont{J.}~\bibnamefont{Wilson}},
  \bibinfo{author}{\bibfnamefont{G.}~\bibnamefont{Mathews}}, \bibnamefont{and}
  \bibinfo{author}{\bibfnamefont{P.}~\bibnamefont{Marronetti}},
  \bibinfo{journal}{Phys.Rev.} \textbf{\bibinfo{volume}{D54}},
  \bibinfo{pages}{1317} (\bibinfo{year}{1996}), \eprint{gr-qc/9601017}.

\bibitem[{\citenamefont{York}(1999)}]{York:1998hy}
\bibinfo{author}{\bibfnamefont{J.}~\bibnamefont{York},
  \bibfnamefont{James~W.}}, \bibinfo{journal}{Phys.Rev.Lett.}
  \textbf{\bibinfo{volume}{82}}, \bibinfo{pages}{1350} (\bibinfo{year}{1999}),
  \eprint{gr-qc/9810051}.

\bibitem[{\citenamefont{Tichy}(2017)}]{Tichy:2016vmv}
\bibinfo{author}{\bibfnamefont{W.}~\bibnamefont{Tichy}},
  \bibinfo{journal}{Rept. Prog. Phys.} \textbf{\bibinfo{volume}{80}},
  \bibinfo{pages}{026901} (\bibinfo{year}{2017}), \eprint{1610.03805}.

\bibitem[{\citenamefont{Bernuzzi and Hilditch}(2010)}]{Bernuzzi:2009ex}
\bibinfo{author}{\bibfnamefont{S.}~\bibnamefont{Bernuzzi}} \bibnamefont{and}
  \bibinfo{author}{\bibfnamefont{D.}~\bibnamefont{Hilditch}},
  \bibinfo{journal}{Phys. Rev.} \textbf{\bibinfo{volume}{D81}},
  \bibinfo{pages}{084003} (\bibinfo{year}{2010}), \eprint{0912.2920}.

\bibitem[{\citenamefont{Hilditch et~al.}(2013)\citenamefont{Hilditch, Bernuzzi,
  Thierfelder, Cao, Tichy et~al.}}]{Hilditch:2012fp}
\bibinfo{author}{\bibfnamefont{D.}~\bibnamefont{Hilditch}},
  \bibinfo{author}{\bibfnamefont{S.}~\bibnamefont{Bernuzzi}},
  \bibinfo{author}{\bibfnamefont{M.}~\bibnamefont{Thierfelder}},
  \bibinfo{author}{\bibfnamefont{Z.}~\bibnamefont{Cao}},
  \bibinfo{author}{\bibfnamefont{W.}~\bibnamefont{Tichy}},
  \bibnamefont{et~al.}, \bibinfo{journal}{Phys. Rev.}
  \textbf{\bibinfo{volume}{D88}}, \bibinfo{pages}{084057}
  (\bibinfo{year}{2013}), \eprint{1212.2901}.

\bibitem[{\citenamefont{Bona et~al.}(1996)\citenamefont{Bona, Mass{\'o}, Stela,
  and Seidel}}]{Bona:1994a}
\bibinfo{author}{\bibfnamefont{C.}~\bibnamefont{Bona}},
  \bibinfo{author}{\bibfnamefont{J.}~\bibnamefont{Mass{\'o}}},
  \bibinfo{author}{\bibfnamefont{J.}~\bibnamefont{Stela}}, \bibnamefont{and}
  \bibinfo{author}{\bibfnamefont{E.}~\bibnamefont{Seidel}}, in
  \emph{\bibinfo{booktitle}{The Seventh {M}arcel {G}rossmann Meeting: On Recent
  Developments in Theoretical and Experimental General Relativity, Gravitation,
  and Relativistic Field Theories}}, edited by
  \bibinfo{editor}{\bibfnamefont{R.~T.} \bibnamefont{Jantzen}},
  \bibinfo{editor}{\bibfnamefont{G.~M.} \bibnamefont{Keiser}},
  \bibnamefont{and} \bibinfo{editor}{\bibfnamefont{R.}~\bibnamefont{Ruffini}}
  (\bibinfo{publisher}{World {S}cientific}, \bibinfo{address}{Singapore},
  \bibinfo{year}{1996}).

\bibitem[{\citenamefont{Alcubierre et~al.}(2003)\citenamefont{Alcubierre,
  Br{\"u}gmann, Diener, Koppitz, Pollney et~al.}}]{Alcubierre:2002kk}
\bibinfo{author}{\bibfnamefont{M.}~\bibnamefont{Alcubierre}},
  \bibinfo{author}{\bibfnamefont{B.}~\bibnamefont{Br{\"u}gmann}},
  \bibinfo{author}{\bibfnamefont{P.}~\bibnamefont{Diener}},
  \bibinfo{author}{\bibfnamefont{M.}~\bibnamefont{Koppitz}},
  \bibinfo{author}{\bibfnamefont{D.}~\bibnamefont{Pollney}},
  \bibnamefont{et~al.}, \bibinfo{journal}{Phys.Rev.}
  \textbf{\bibinfo{volume}{D67}}, \bibinfo{pages}{084023}
  (\bibinfo{year}{2003}), \eprint{gr-qc/0206072}.

\bibitem[{\citenamefont{van Meter et~al.}(2006)\citenamefont{van Meter, Baker,
  Koppitz, and Choi}}]{vanMeter:2006vi}
\bibinfo{author}{\bibfnamefont{J.~R.} \bibnamefont{van Meter}},
  \bibinfo{author}{\bibfnamefont{J.~G.} \bibnamefont{Baker}},
  \bibinfo{author}{\bibfnamefont{M.}~\bibnamefont{Koppitz}}, \bibnamefont{and}
  \bibinfo{author}{\bibfnamefont{D.-I.} \bibnamefont{Choi}},
  \bibinfo{journal}{Phys. Rev.} \textbf{\bibinfo{volume}{D73}},
  \bibinfo{pages}{124011} (\bibinfo{year}{2006}), \eprint{gr-qc/0605030}.

\bibitem[{\citenamefont{Borges et~al.}(2008)\citenamefont{Borges, Carmona,
  Costa, and Don}}]{Borges:2008a}
\bibinfo{author}{\bibfnamefont{R.}~\bibnamefont{Borges}},
  \bibinfo{author}{\bibfnamefont{M.}~\bibnamefont{Carmona}},
  \bibinfo{author}{\bibfnamefont{B.}~\bibnamefont{Costa}}, \bibnamefont{and}
  \bibinfo{author}{\bibfnamefont{W.~S.} \bibnamefont{Don}},
  \bibinfo{journal}{Journal of Computational Physics}
  \textbf{\bibinfo{volume}{227}}, \bibinfo{pages}{3191} (\bibinfo{year}{2008}),
  ISSN \bibinfo{issn}{0021-9991},
  \urlprefix\url{http://www.sciencedirect.com/science/article/pii/S0021999107005232}.

\bibitem[{\citenamefont{Bernuzzi and Dietrich}(2016)}]{Bernuzzi:2016pie}
\bibinfo{author}{\bibfnamefont{S.}~\bibnamefont{Bernuzzi}} \bibnamefont{and}
  \bibinfo{author}{\bibfnamefont{T.}~\bibnamefont{Dietrich}},
  \bibinfo{journal}{Phys. Rev.} \textbf{\bibinfo{volume}{D94}},
  \bibinfo{pages}{064062} (\bibinfo{year}{2016}), \eprint{1604.07999}.

\bibitem[{\citenamefont{Dietrich
  et~al.}(2018{\natexlab{a}})\citenamefont{Dietrich, Bernuzzi, Bruegmann, and
  Tichy}}]{Dietrich:2018upm}
\bibinfo{author}{\bibfnamefont{T.}~\bibnamefont{Dietrich}},
  \bibinfo{author}{\bibfnamefont{S.}~\bibnamefont{Bernuzzi}},
  \bibinfo{author}{\bibfnamefont{B.}~\bibnamefont{Bruegmann}},
  \bibnamefont{and} \bibinfo{author}{\bibfnamefont{W.}~\bibnamefont{Tichy}}, in
  \emph{\bibinfo{booktitle}{{Proceedings, 26th Euromicro International
  Conference on Parallel, Distributed and Network-based Processing (PDP 2018):
  Cambridge, UK, March 21-23, 2018}}} (\bibinfo{year}{2018}{\natexlab{a}}), pp.
  \bibinfo{pages}{682--689}, \eprint{1803.07965}.

\bibitem[{\citenamefont{Dudi et~al.}(2018)\citenamefont{Dudi, Pannarale,
  Dietrich, Hannam, Bernuzzi, Ohme, and Brügmann}}]{Dudi:2018jzn}
\bibinfo{author}{\bibfnamefont{R.}~\bibnamefont{Dudi}},
  \bibinfo{author}{\bibfnamefont{F.}~\bibnamefont{Pannarale}},
  \bibinfo{author}{\bibfnamefont{T.}~\bibnamefont{Dietrich}},
  \bibinfo{author}{\bibfnamefont{M.}~\bibnamefont{Hannam}},
  \bibinfo{author}{\bibfnamefont{S.}~\bibnamefont{Bernuzzi}},
  \bibinfo{author}{\bibfnamefont{F.}~\bibnamefont{Ohme}}, \bibnamefont{and}
  \bibinfo{author}{\bibfnamefont{B.}~\bibnamefont{Brügmann}},
  \bibinfo{journal}{Phys. Rev.} \textbf{\bibinfo{volume}{D98}},
  \bibinfo{pages}{084061} (\bibinfo{year}{2018}), \eprint{1808.09749}.

\bibitem[{\citenamefont{Tsang et~al.}(2018)\citenamefont{Tsang, Rollier, Ghosh,
  Samajdar, Agathos, Chatziioannou, Cardoso, Khanna, and Van
  Den~Broeck}}]{Tsang:2018uie}
\bibinfo{author}{\bibfnamefont{K.~W.} \bibnamefont{Tsang}},
  \bibinfo{author}{\bibfnamefont{M.}~\bibnamefont{Rollier}},
  \bibinfo{author}{\bibfnamefont{A.}~\bibnamefont{Ghosh}},
  \bibinfo{author}{\bibfnamefont{A.}~\bibnamefont{Samajdar}},
  \bibinfo{author}{\bibfnamefont{M.}~\bibnamefont{Agathos}},
  \bibinfo{author}{\bibfnamefont{K.}~\bibnamefont{Chatziioannou}},
  \bibinfo{author}{\bibfnamefont{V.}~\bibnamefont{Cardoso}},
  \bibinfo{author}{\bibfnamefont{G.}~\bibnamefont{Khanna}}, \bibnamefont{and}
  \bibinfo{author}{\bibfnamefont{C.}~\bibnamefont{Van Den~Broeck}},
  \bibinfo{journal}{Phys. Rev.} \textbf{\bibinfo{volume}{D98}},
  \bibinfo{pages}{024023} (\bibinfo{year}{2018}), \eprint{1804.04877}.

\bibitem[{\citenamefont{Chatziioannou et~al.}(2017)\citenamefont{Chatziioannou,
  Clark, Bauswein, Millhouse, Littenberg, and Cornish}}]{Chatziioannou:2017ixj}
\bibinfo{author}{\bibfnamefont{K.}~\bibnamefont{Chatziioannou}},
  \bibinfo{author}{\bibfnamefont{J.~A.} \bibnamefont{Clark}},
  \bibinfo{author}{\bibfnamefont{A.}~\bibnamefont{Bauswein}},
  \bibinfo{author}{\bibfnamefont{M.}~\bibnamefont{Millhouse}},
  \bibinfo{author}{\bibfnamefont{T.~B.} \bibnamefont{Littenberg}},
  \bibnamefont{and} \bibinfo{author}{\bibfnamefont{N.}~\bibnamefont{Cornish}},
  \bibinfo{journal}{Phys. Rev.} \textbf{\bibinfo{volume}{D96}},
  \bibinfo{pages}{124035} (\bibinfo{year}{2017}), \eprint{1711.00040}.

\bibitem[{\citenamefont{Metzger}(2017)}]{Metzger:2016pju}
\bibinfo{author}{\bibfnamefont{B.~D.} \bibnamefont{Metzger}},
  \bibinfo{journal}{Living Rev. Rel.} \textbf{\bibinfo{volume}{20}},
  \bibinfo{pages}{3} (\bibinfo{year}{2017}), \eprint{1610.09381}.

\bibitem[{\citenamefont{Tanaka}(2016)}]{Tanaka:2016sbx}
\bibinfo{author}{\bibfnamefont{M.}~\bibnamefont{Tanaka}},
  \bibinfo{journal}{Adv. Astron.} \textbf{\bibinfo{volume}{2016}},
  \bibinfo{pages}{6341974} (\bibinfo{year}{2016}), \eprint{1605.07235}.

\bibitem[{\citenamefont{Shibata et~al.}(2017)\citenamefont{Shibata,
  Fujibayashi, Hotokezaka, Kiuchi, Kyutoku, Sekiguchi, and
  Tanaka}}]{Shibata:2017xdx}
\bibinfo{author}{\bibfnamefont{M.}~\bibnamefont{Shibata}},
  \bibinfo{author}{\bibfnamefont{S.}~\bibnamefont{Fujibayashi}},
  \bibinfo{author}{\bibfnamefont{K.}~\bibnamefont{Hotokezaka}},
  \bibinfo{author}{\bibfnamefont{K.}~\bibnamefont{Kiuchi}},
  \bibinfo{author}{\bibfnamefont{K.}~\bibnamefont{Kyutoku}},
  \bibinfo{author}{\bibfnamefont{Y.}~\bibnamefont{Sekiguchi}},
  \bibnamefont{and} \bibinfo{author}{\bibfnamefont{M.}~\bibnamefont{Tanaka}},
  \bibinfo{journal}{Phys. Rev.} \textbf{\bibinfo{volume}{D96}},
  \bibinfo{pages}{123012} (\bibinfo{year}{2017}), \eprint{1710.07579}.

\bibitem[{\citenamefont{Radice et~al.}(2018)\citenamefont{Radice, Perego,
  Hotokezaka, Fromm, Bernuzzi, and Roberts}}]{Radice:2018pdn}
\bibinfo{author}{\bibfnamefont{D.}~\bibnamefont{Radice}},
  \bibinfo{author}{\bibfnamefont{A.}~\bibnamefont{Perego}},
  \bibinfo{author}{\bibfnamefont{K.}~\bibnamefont{Hotokezaka}},
  \bibinfo{author}{\bibfnamefont{S.~A.} \bibnamefont{Fromm}},
  \bibinfo{author}{\bibfnamefont{S.}~\bibnamefont{Bernuzzi}}, \bibnamefont{and}
  \bibinfo{author}{\bibfnamefont{L.~F.} \bibnamefont{Roberts}},
  \bibinfo{journal}{Astrophys. J.} \textbf{\bibinfo{volume}{869}},
  \bibinfo{pages}{130} (\bibinfo{year}{2018}), \eprint{1809.11161}.

\bibitem[{\citenamefont{Iwazaki}(2015)}]{Iwazaki:2014wka}
\bibinfo{author}{\bibfnamefont{A.}~\bibnamefont{Iwazaki}},
  \bibinfo{journal}{Phys. Rev.} \textbf{\bibinfo{volume}{D91}},
  \bibinfo{pages}{023008} (\bibinfo{year}{2015}), \eprint{1410.4323}.

\bibitem[{\citenamefont{Bai and Hamada}(2018)}]{Bai:2017feq}
\bibinfo{author}{\bibfnamefont{Y.}~\bibnamefont{Bai}} \bibnamefont{and}
  \bibinfo{author}{\bibfnamefont{Y.}~\bibnamefont{Hamada}},
  \bibinfo{journal}{Phys. Lett.} \textbf{\bibinfo{volume}{B781}},
  \bibinfo{pages}{187} (\bibinfo{year}{2018}), \eprint{1709.10516}.

\bibitem[{\citenamefont{Dietrich
  et~al.}(2018{\natexlab{b}})\citenamefont{Dietrich, Day, Clough, Coughlin, and
  Niemeyer}}]{Dietrich:2018jov}
\bibinfo{author}{\bibfnamefont{T.}~\bibnamefont{Dietrich}},
  \bibinfo{author}{\bibfnamefont{F.}~\bibnamefont{Day}},
  \bibinfo{author}{\bibfnamefont{K.}~\bibnamefont{Clough}},
  \bibinfo{author}{\bibfnamefont{M.}~\bibnamefont{Coughlin}}, \bibnamefont{and}
  \bibinfo{author}{\bibfnamefont{J.}~\bibnamefont{Niemeyer}},
  \bibinfo{journal}{arXiv: 1808.04746}  (\bibinfo{year}{2018}{\natexlab{b}}),
  \eprint{1808.04746}.

\bibitem[{\citenamefont{Capano et~al.}(2019)\citenamefont{Capano, Tews, Brown,
  Margalit, De, Kumar, Brown, Krishnan, and Reddy}}]{Capano:2019eae}
\bibinfo{author}{\bibfnamefont{C.~D.} \bibnamefont{Capano}},
  \bibinfo{author}{\bibfnamefont{I.}~\bibnamefont{Tews}},
  \bibinfo{author}{\bibfnamefont{S.~M.} \bibnamefont{Brown}},
  \bibinfo{author}{\bibfnamefont{B.}~\bibnamefont{Margalit}},
  \bibinfo{author}{\bibfnamefont{S.}~\bibnamefont{De}},
  \bibinfo{author}{\bibfnamefont{S.}~\bibnamefont{Kumar}},
  \bibinfo{author}{\bibfnamefont{D.~A.} \bibnamefont{Brown}},
  \bibinfo{author}{\bibfnamefont{B.}~\bibnamefont{Krishnan}}, \bibnamefont{and}
  \bibinfo{author}{\bibfnamefont{S.}~\bibnamefont{Reddy}}
  (\bibinfo{year}{2019}), \eprint{1908.10352}.

\end{thebibliography}

\end{document}